\documentclass{article}

\PassOptionsToPackage{numbers, compress}{natbib}

\usepackage[table]{xcolor}
\usepackage{subcaption} 
\usepackage[main, final]{neurips_2026}

\PassOptionsToPackage{numbers,compress}{natbib}
\usepackage[table]{xcolor}
\usepackage{subcaption}
\usepackage{neurips_2026}
\usepackage[utf8]{inputenc}
\usepackage[T1]{fontenc}
\usepackage{hyperref}
\usepackage{url}
\usepackage{amsfonts}
\usepackage{amsmath}
\usepackage{nicefrac}
\usepackage{microtype}
\usepackage{makecell}
\usepackage{graphicx}
\usepackage{tcolorbox}
\tcbuselibrary{skins,breakable}
\usepackage{multirow}
\usepackage{tikz}
\usepackage{booktabs}
\usepackage{colortbl}
\usepackage{etoolbox}
\usepackage{array}
\usepackage{threeparttable}
\usepackage{xfp}
\usepackage{longtable}
\usepackage{wrapfig}
\usepackage[table]{xcolor}
\usepackage{makecell}
\usepackage{colortbl}
\usepackage{cleveref}

\usepackage{algorithm}
\usepackage{algpseudocode}

\definecolor{familybg}{gray}{0.88}
\definecolor{headergreen}{RGB}{198,224,180}
\definecolor{headerblue}{RGB}{189,215,238}

\definecolor{heatA}{HTML}{FFF7BC} 
\definecolor{heatB}{HTML}{FEE391}
\definecolor{heatC}{HTML}{FED976}
\definecolor{heatD}{HTML}{FEB24C}
\definecolor{heatE}{HTML}{FD8D3C}
\definecolor{heatF}{HTML}{D94801} 

\newcolumntype{C}[1]{>{\centering\arraybackslash}p{#1}}
\newcolumntype{L}[1]{>{\raggedright\arraybackslash}p{#1}}

\newcommand{\modelcell}[1]{#1}

\newcommand{\heatbg}[1]{%
  \ifdim #1pt < 60pt heatA\else
    \ifdim #1pt < 70pt heatB\else
      \ifdim #1pt < 78pt heatC\else
        \ifdim #1pt < 85pt heatD\else
          \ifdim #1pt < 92pt heatE\else
            heatF%
          \fi
        \fi
      \fi
    \fi
  \fi
}

\renewcommand{\arraystretch}{1.06}

\newcommand{\heatcellscore}[1]{%
  \cellcolor{\heatbg{#1}}%
  \parbox[c][2.35em][c]{\linewidth}{%
    \centering
    {\fontsize{13}{14}\selectfont #1\par}%
  }%
}

\newtcolorbox[auto counter]{promptbox}[1]{
    enhanced,
    breakable,
    colback=gray!4,
    colframe=black!55,
    boxrule=0.6pt,
    arc=1mm,
    left=2mm,
    right=2mm,
    top=1.5mm,
    bottom=1.5mm,
    before skip=6pt,
    after skip=6pt,
    fonttitle=\bfseries,
    title={Box~\thetcbcounter: #1},
}

\title{Known By Their Actions: Fingerprinting LLM Browser Agents via UI Traces}

\author{%
  William Lugoloobi$^{1}$\thanks{Kindly correspond to \texttt{william.lugoloobi@oii.ox.ac.uk}}\quad
  Samuele Marro$^{2,3}$ \quad
  Jabez Magomere$^{1}$ \quad
  Joss Wright$^{1}$ \quad
  Chris Russell$^{1}$ \\
  $^{1}$Oxford Internet Institute, University of Oxford \\
  $^{2}$Department of Engineering Science, University of Oxford \\
  $^{3}$Institute for Decentralized AI \\
  }

\begin{document}

\maketitle

\begin{abstract}
As LLM-based agents increasingly browse the web on users' behalf, a natural question arises: can websites passively identify which underlying model powers an agent? Doing so would represent a significant security risk, enabling targeted attacks tailored to known model vulnerabilities. Across 14 frontier LLMs and four web environments spanning information retrieval and shopping tasks, we show that an agent's actions and interaction timings, captured via a passive JavaScript tracker, are sufficient to identify the underlying model with up to 96\% F1 within a fixed harness. Although fingerprints are harness-specific, with cross-harness transfer falling to near chance, pooling traces across harnesses recovers model identification at 0.74 to 0.81 Macro-F1. We further show that strong classifiers can be trained from roughly one third of the available traces, and that identification improves steadily as more events are observed. Injecting randomised timing delays between actions substantially degrades classifier performance, but does not provide robust protection: a classifier retrained on delayed traces largely recovers performance. We release our code and a labelled corpus of agent traces at \nolinkurl{https://github.com/KabakaWilliam/known_actions}.
\end{abstract}

\section{Introduction}
LLM-based agents that browse the web and operate computer 
interfaces on behalf of users are moving rapidly from research 
prototypes to production \citep{nakano_webgpt_2022, 
yao_react_2022, wang_survey_2024}. As these systems are deployed at scale across live websites, every page they visit becomes a potential observation point, and we show that observation alone is enough to identify the model. This exposes users to potential security risks.
\begin{figure}[h!]
    \centering
    \includegraphics[width=\linewidth]{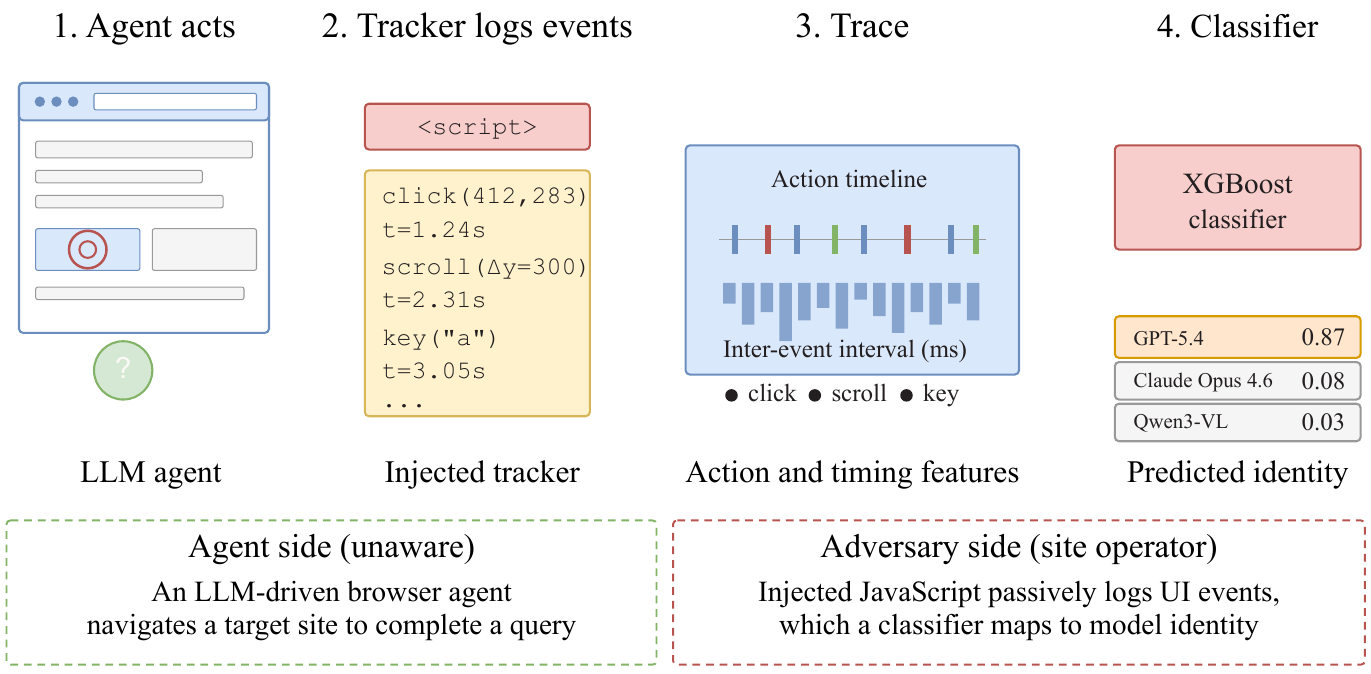}
    \caption{\textbf{Overview of our trace collection and threat model.} Using a JavaScript tracker, we collect actions performed by a browsing agent and train a classifier to predict model identity.}
    \label{fig:hero}
\end{figure}

Every agent visit to a website leaves a trace of clicks, scrolls, 
keypresses, and other \textit{actions} observable to any party 
controlling the page. Prior work has established that behavioural 
traces like this distinguish human users from automated clients 
\citep{iliou_detection_2021, acien_becaptcha-mouse_2022}, and that 
passively collected browser attributes re-identify human users 
across sessions \citep{eckersley_how_2010, vastel_fp-stalker_2018}. 
These works ask a binary question: human or bot. As LLM-based 
agents displace scripted automation, we probe deeper: \textit{given  that the client is an agent, which model is pulling the strings?} 
In classical security frameworks, target identification is the 
first step toward exploitation~\citep{hutchins_intelligence-driven_2011}. 
Thus for LLM-based agents, knowing the underlying model enables more targeted attacks: an adversary can select from a known 
set of model-specific jailbreaks or reduce the search space for a 
white-box adversarial attack~\citep{pasquini_llmmap_2025}.

To our knowledge, we are the first to show that the underlying foundation model of a browser agent can be inferred from passive in-page UI traces alone. Concurrent work \citep{wang_fp-agent_2026} shows that AI browsing agents can be distinguished from human visitors and from each other using behavioural and network features. Its classes, however, are agent products (e.g., ChatGPT Atlas, Comet, Browser Use), each running its own harness, and we find that changing the harness alone is enough to break a model classifier (see \Cref{app:harness-transfer}). We address the complementary, subsequent problem of attributing a session already believed to be an agent to a specific model, rather than its harness.
Using lightweight classifiers trained on UI action traces 
collected via injected JavaScript, we achieve agent identification 
F1 scores of up to 96\% across 14 frontier LLMs. The identification does not rely on browser attributes and headers (which can be spoofed), but on the temporal and 
structural dynamics of how different models navigate, click, and 
interact with page elements. In other words, the behaviour of a model is sufficient to accurately identify it. For adversaries, this means that agents can be identified and exploited.

The main contributions of this paper are as follows:
\begin{itemize}
\item \textbf{Agent actions are a fingerprint of model identity.}
We demonstrate for the first time that the on-page actions 
of LLM browser agents encodes the identity of the underlying model, 
achieving up to 96\% classification Macro F1 across 14 frontier 
models using only behavioural traces collected via passive 
JavaScript injection.

\item \textbf{A formalised threat model with defence analysis.} We characterise agent fingerprinting under a passive, co-located adversary and show that the fingerprinter is fast and cheap to update once labelled traces of a new model--harness pair are available. We discuss this more in Appendix~\ref{app:scaling}. We also test two natural defences, randomised timing delays and prompt normalisation. Both weaken a classifier trained on unperturbed traces, but neither holds up once the adversary retrains on matching traces.

\item \textbf{Resources for agent fingerprinting.} We release a labelled corpus of agent interaction traces across four web environments and a browser harness compatible with both closed and open-source LLMs, enabling reproducible research into behavioural attribution of LLM agents.

\end{itemize}


\section{Related Work}

\textbf{LLM-based web agents.} Autonomous agents that combine 
language models with browser automation have rapidly moved from 
research prototypes to production systems \citep{nakano_webgpt_2022, 
yao_react_2022, wang_survey_2024}. Benchmarks such as WebArena 
\citep{zhou_webarena_2023} and Mind2Web \citep{deng_mind2web_2023} 
have established standard evaluation environments for these systems. 
As these agents are deployed at scale, the question of whether 
a site operator can determine which model is visiting becomes 
practically consequential for access control, content delivery, 
and adversarial exploitation.

\textbf{Bot detection and browser fingerprinting.}
A large body of work studies how to distinguish automated from human web traffic.
Early approaches relied on request-pattern heuristics and crawl detection
\citep{kabe_determining_2000, huntington_web_2008, geens_evaluation_2006, koster_web_1994,koster_robots_2022, doran_web_2011}.
Subsequent work has shown that fine-grained behavioural signals, such as mouse
movements and interaction timing encode neuromotor structure that can reliably
separate humans from bots \citep{fu_rumba-mouse_2020,wang_optimizing_2025, acien_becaptcha-mouse_2022, chong_user_2020}.
More recent systems combine behavioural features with server-side logs to detect
increasingly sophisticated bots that mimic realistic browser fingerprints
\citep{iliou_detection_2021}.

In parallel, browser fingerprinting research has demonstrated that passively
collected client-side attributes (e.g., canvas, fonts, WebGL) can be combined
into persistent identifiers at scale \citep{eckersley_how_2010, vastel_fp-stalker_2018}.
Across these lines of work, the problem is typically framed as binary
classification: human versus bot.
We instead consider a finer-grained setting: given that the user is an
LLM-based agent, can we identify \emph{which model} produced the interaction trace solely from its actions?

\textbf{Side-channel attacks and traffic fingerprinting.} 
Side-channel attacks show that systems leak sensitive information 
through correlated observables, even when their outputs do not 
\citep{kocher_timing_1996}. Applied to web traffic, deep learning 
classifiers trained on encrypted packet sequences identify visited 
websites with over 98\% accuracy \citep{rimmer_automated_2018, 
sirinam_deep_2018}. \citet{cook_theres_2022} draw a 
useful distinction between on-path attackers, who observe network 
traffic from a separate machine, and co-located attackers, whose 
code runs on the same machine as the victim. Our attack is 
co-located: we inject JavaScript trackers into the page and collect 
action traces directly, without any network-level visibility.

\textbf{LLM and agent fingerprinting.} \citet{pasquini_llmmap_2025} 
fingerprint LLM-integrated applications by sending crafted queries 
and analysing responses, achieving over 95\% accuracy across 42 
model versions. Beyond the result, they demonstrate that knowing the 
underlying model enables targeted attacks: an adversary who can identify 
the model can craft inputs that exploit model-specific behaviours, biases, 
or known failure modes, turning fingerprinting from a reconnaissance step 
\citep{hutchins_intelligence-driven_2011, mitre_corporation_mitre_2025, 
zychlinski_whole_2025} into an attack primitive.  Closest to our setting, \citet{zhang_exposing_2025} show that 
task-specific LLM agent applications leave distinct network traffic 
fingerprints. Their attack observes packet-level metadata generated by 
agent tool use, and uses it to infer behaviours, application identity, and downstream user attributes. Concurrently, \citet{wang_fp-agent_2026} classify seven commercial and open-source browsing agents from in-page behavioural and browser-fingerprint features.

In contrast, we study model attribution from ordinary UI events generated while an agent browses a website. Because all agents in a given experiment share the same browser harness and action space, our classifier separates models within that setting rather than distinguishing applications or interaction patterns. The resulting fingerprint is tied to the model--harness pair (Appendix~\ref{app:harness-transfer}), but our results show that the attribution surface extends beyond network observers and active probers to the visited page itself.

\section{Problem Formulation}\label{sec:formulation}

\subsection{Agent Identification as a Classification Problem}

We study whether interaction traces produced by web-browsing agents 
contain sufficient signal to identify the underlying language model, 
and formalise this as a supervised classification problem over 
behavioural traces.

\paragraph{Agent and environment.}
An agent $a \in \mathcal{A}$ is instantiated by a language model 
$m \in \mathcal{M}$ interacting with a web environment $\mathcal{E}$ 
through a fixed browser harness $h$. At each timestep $t$, the agent 
conditions on the current observation $s_t$ (a rendered screenshot), 
updates an internal plan, and produces an action $u_t$ (e.g., click, 
scroll, keypress). The environment executes $u_t$, yielding a new 
observation $s_{t+1}$, and the process repeats for $N$ planning 
steps.

We assume all agents share the same interface and action space. Within an experiment this removes the harness as a source of variation, but it does not make the fingerprint harness-independent: the trace reflects the model--harness pairing. We test this with a second harness in Appendix~\ref{app:harness-transfer}. Throughout, identifying a model means identifying it as driven through a given harness.

\paragraph{Interaction trace.}
A session generates a trace
\[
\tau = \{(s_t, u_t, \Delta t_t)\}_{t=1}^{T},
\]
where $\Delta t_t$ denotes the time elapsed between consecutive 
actions. We restrict our analysis to client-side interaction 
signals, that is, the sequence and timing of actions produced by 
the agent, and do not use server-side metadata such as headers, 
IP addresses, or TLS fingerprints. Traces may span multiple pages 
within the same host and are treated as belonging to a single 
session tied to a query $Q$. We further restrict traces to a 
single domain to control for environmental variability.

\paragraph{Identification task.}
Given a trace $\tau$ generated by an unknown agent, the goal is 
to predict the originating model
\[
\hat{m} = f(\tau), \quad \hat{m} \in \mathcal{M},
\]
where $f$ is a classifier trained on labelled traces. Unlike prior 
work that frames this as human versus bot discrimination, we assume the client is automated and ask which model produced the behaviour.

\paragraph{Feature creation.}
We consider feature mappings $\phi(\tau)$ derived from Inter-Event Intervals (IEIs, the time between two consecutive actions), navigation structure 
(e.g., click frequency and page transitions), and interaction 
patterns (e.g., action type distributions). These features are 
designed to capture behavioural regularities induced by the 
underlying model. Full descriptions of extracted features are 
provided in Appendix~\ref{app:behavioral_features}.

\subsection{Threat Model}

We consider a passive, co-located adversary: a site operator who injects lightweight JavaScript into pages to collect actions performed by an agent while visiting. Additionally, the adversary is assumed to have already established that the session originates from an automated agent rather than a human user, using classical bot-detection methods \citep{iliou_detection_2021} or detectors designed for AI browsing agents \citep{wang_fp-agent_2026}. Consequently, the identification problem becomes one of model attribution, not human-versus-bot detection; detecting such sessions is a separate problem.

The adversary has access to the sequence and timing of on-page 
actions, but no access to model internals, generated text, or 
network-layer traffic. We assume a realistic setting in which 
standard browser fingerprint signals are present alongside 
behavioural traces, and a fresh browser is instantiated for each 
session. The adversary is passive by assumption: they cannot 
modify page content or craft adversarial inputs to probe the 
agent directly. The adversary's objective is identification: 
once the underlying model is known, they can consult a 
corpus of model-specific jailbreaks~\citep{zou_universal_2023, 
shen_anything_2024} or initialise a targeted optimisation 
procedure~\citep{pasquini_llmmap_2025, aichberger_attacking_2025, seip_preference_2026} with a substantially 
reduced search space, bypassing the cost of generic black-box probing entirely.

Depending on the adversary's knowledge of the agent population, this identification problem takes two forms.

\paragraph{Closed-set fingerprinting.}
Let $\mathcal{A} = \{a_1, \dots, a_K\}$ be the full set of $K$ agents. The 
adversary assumes that any observed trace originates from some $a_i \in \mathcal{A}$ 
and learns a classifier $f: \mathcal{T} \rightarrow \mathcal{A}$ over all $K$ 
classes, where $\mathcal{T}$ denotes the space of interaction traces $\tau$. At evaluation time, test traces are drawn from the same set 
$\mathcal{A}$, so the problem reduces to standard multi-class classification.

Notably, a closed-set classifier can be \textbf{cheaply updated as new models are released}: using the same collection pipeline, an adversary can gather traces from a new model--harness pairing and retrain the classifier in seconds (Table~\ref{app:tab:scaling}).

\paragraph{Open-set fingerprinting.}
In a more realistic setting, the adversary cannot know every agent they may
encounter. Let $\mathcal{A}_{\text{train}} \subset \mathcal{A}$ be the agents
known at training time, and let $\mathcal{A}_{\text{unk}} = \mathcal{A} \setminus
\mathcal{A}_{\text{train}}$ denote the unknown agents. The classifier must either
assign a trace to a known agent class or flag it as \textit{unknown}.

We instantiate this setting via a \textbf{leave-one-agent-out (LOO)} protocol.
Let $K = |\mathcal{A}|$. For each held-out agent $a_i \in \mathcal{A}$, we train
a classifier on traces from $\mathcal{A} \setminus \{a_i\}$ (all agents except $a_i$) and evaluate on the
test-split traces of the $K-1$ known agents together with all traces from $a_i$
as the unknown class. We measure the ability to separate known from unknown traces
using AUROC over the binary known/unknown discrimination, reported separately for
each held-out agent $a_i$, yielding $K$ values across the full agent set.





\section{Experimental Setup}\label{sec:experimentalSetup}

\paragraph{Data}

We construct a dataset spanning two broad task domains where agents have been widely applied: information seeking tasks and online shopping. For question answering, we repurpose 2WikiMultiHop \citep{ho_constructing_2020} and FRAMES \citep{krishna_fact_2025} as live web tasks, which requires models to navigate and retrieve information across multiple pages. Similarly, for shopping, we adapt the e-commerce benchmarks Webshop \citep{yao_webshop_2023} and Deepshop \citep{lyu_deepshop_2025}. Standard train, validation, and test splits for all datasets are summarised in Table~\ref{tab:Dataset_Breakdown}. Together, these environments provide a broad basis for eliciting and comparing behavioural fingerprints across diverse interaction regimes. This structure also lets us distinguish in-domain attribution, cross-task transfer within a website, pooled site-level training, and cross-site transfer; we report these generalisation experiments in Appendix~\ref{app:site_transfer}.

\begin{table}[h!]
\centering
\caption{Dataset splits used in this work. All benchmarks are deployed as live 
web tasks on their respective target websites.}
\label{tab:Dataset_Breakdown}
\small
\begin{tabular}{llccccc}
\toprule
\textbf{Dataset} & \textbf{Domain} & \textbf{Target Website} & \textbf{Train} & \textbf{Val} & \textbf{Test} & \textbf{Total} \\
\midrule
2WikiMultiHop \citep{ho_constructing_2020} & QA         & Wikipedia.com & 150 & 75  & 75  & 300  \\
FRAMES \citep{krishna_fact_2025} & QA         & Wikipedia.com & 150 & 75  & 75  & 300  \\
Webshop \citep{yao_webshop_2023}           & Shopping   & Amazon.com    & 150 & 75  & 75  & 300  \\
Deepshop \citep{lyu_deepshop_2025}         & Shopping   & Amazon.com    & 75  & 37  & 38  & 150  \\
\midrule
\textbf{Total}                             &            &               & 525 & 262 & 263 & 1050 \\
\bottomrule
\end{tabular}
\end{table}

\paragraph{Models}
We evaluate 14 multimodal LLMs selected to support model identity classification at two levels of granularity: model family (e.g., Qwen3-VL, Qwen3.5-VL) and specific model variant (e.g., Qwen3.5-9B vs.\ Qwen3.5-27B). Full details are given in Appendix~\ref{app:sec:model_details}. Locally hosted open-weight models include: the GLM-4.6V \citep{team_glm-45v_2025}, Qwen3-VL series \citep{team_qwen3_2025}, Qwen3.5-VL series \citep{qwen_team_qwen35_2026}, UI-TARS-1.5-7B \citep{qin_ui-tars_2025} (a UI-specialist fine-tune of Qwen2.5-VL \citep{bai_qwen25-vl_2025}), and the Gemma-4 series \citep{gemma_team_gemma_2026}. We additionally include Seed-2.0-Lite \citep{bytedance_seed_team_seed20_2026}, an open-weight model accessed via OpenRouter. Proprietary frontier models, namely GPT-5.4 \citep{openai_introducing_2026}, Gemini-3.1 and Gemini-3-Flash \citep{google_gemini_nodate}, and Claude Opus 4.6 \citep{anthropic_claude_nodate}, are evaluated via their respective APIs.

\paragraph{Agent Harness}
We standardise our computer-use harness with Midscene.js 
\citep{xiao_zhou_midscenejs_2025}, a JavaScript library that 
provides a standardised interface between multimodal LLMs and 
browser environments, enabling models to perceive and interact 
with web-based UIs by translating actions into Playwright commands. 
All agents share an identical harness configuration, ensuring 
that behavioural differences within an experiment are attributable to 
the underlying model rather than the harness. Since Midscene.js operates in pure-vision mode only, browser observations are limited to visual screenshots; this also suits our setting, as it ensures any identifying signal derives from visual reasoning and interaction behaviour rather than from differences in how models process structured markup. In Appendix~\ref{app:harness-transfer}, we replicate within-harness attribution with a second harness, Browser Use \citep{muller_browser_2024}, which gives the model both screenshots and HTML text. We find that fingerprints are strongly harness-conditioned, but that training on a mix of both harnesses recovers most of the performance.

\paragraph{Trace collection}
We instrument each page with a lightweight JavaScript observer 
injected at session initialisation. The observer attaches 
event listeners to the DOM and records every interaction event 
produced by the agent, including click coordinates and target 
element type, scroll direction and magnitude, keypress events 
and inter-keystroke timing, and navigation events with 
timestamps. All events are logged with millisecond-resolution 
timestamps relative to session start, yielding a raw event 
stream that is post-processed into the structured trace format 
$\tau = \{(s_t, u_t, \Delta t_t)\}_{t=1}^{T}$ defined in 
Section~\ref{sec:formulation}. A fresh browser context is 
instantiated for each session to prevent cross-session state 
leakage.  Each agent completes every 
query in the dataset independently, yielding a labelled corpus 
of traces with model identity as the class label.

\paragraph{Classifiers}
We train five classifier families on the collected traces: Lasso Regression, Logistic Regression, Random Forest, XGBoost, and an LSTM network. We report results primarily for XGBoost, which achieves the strongest performance across datasets; full results for all classifiers are provided in Appendix~\ref{app:classifier_hyperparams}.

\paragraph{Metrics}
To evaluate our classifiers in the closed-set fingerprinting setting, we report the per-LLM F1 score and the macro F1 across all classifiers. For the open-set fingerprinting setting, we report the AUROC for each classifier.

\paragraph{Hardware}
All open-source models are served via vLLM on a node equipped with two NVIDIA
H100 GPUs. We additionally evaluate transfer between vLLM and SGLang in Appendix~\ref{app:engine-transfer}.

\begin{figure}[t]
    \centering
    \includegraphics[width=\linewidth]{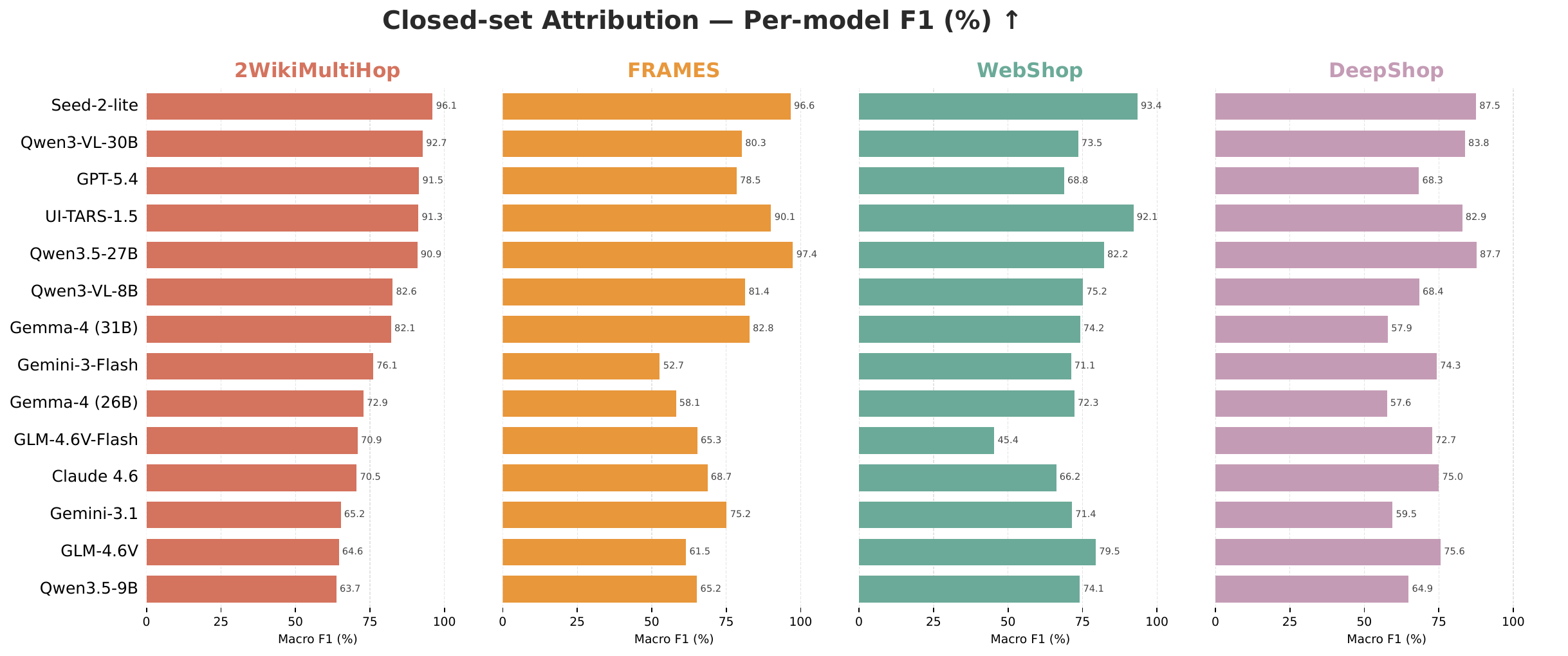}
   \caption{\textbf{Agents are identifiable from traces of their actions.} Across both information-seeking and shopping benchmarks, our XGBoost classifier trained on action traces reliably identifies the browsing agent. High scores across all four datasets indicate that agents leave consistent, distinctive traces in their browsing actions.}
    \label{fig:main_closed_bars}
\end{figure}

\section{Fingerprinting Results}\label{sec:results}

\begin{figure}[t]
    \centering
    \includegraphics[width=\linewidth]{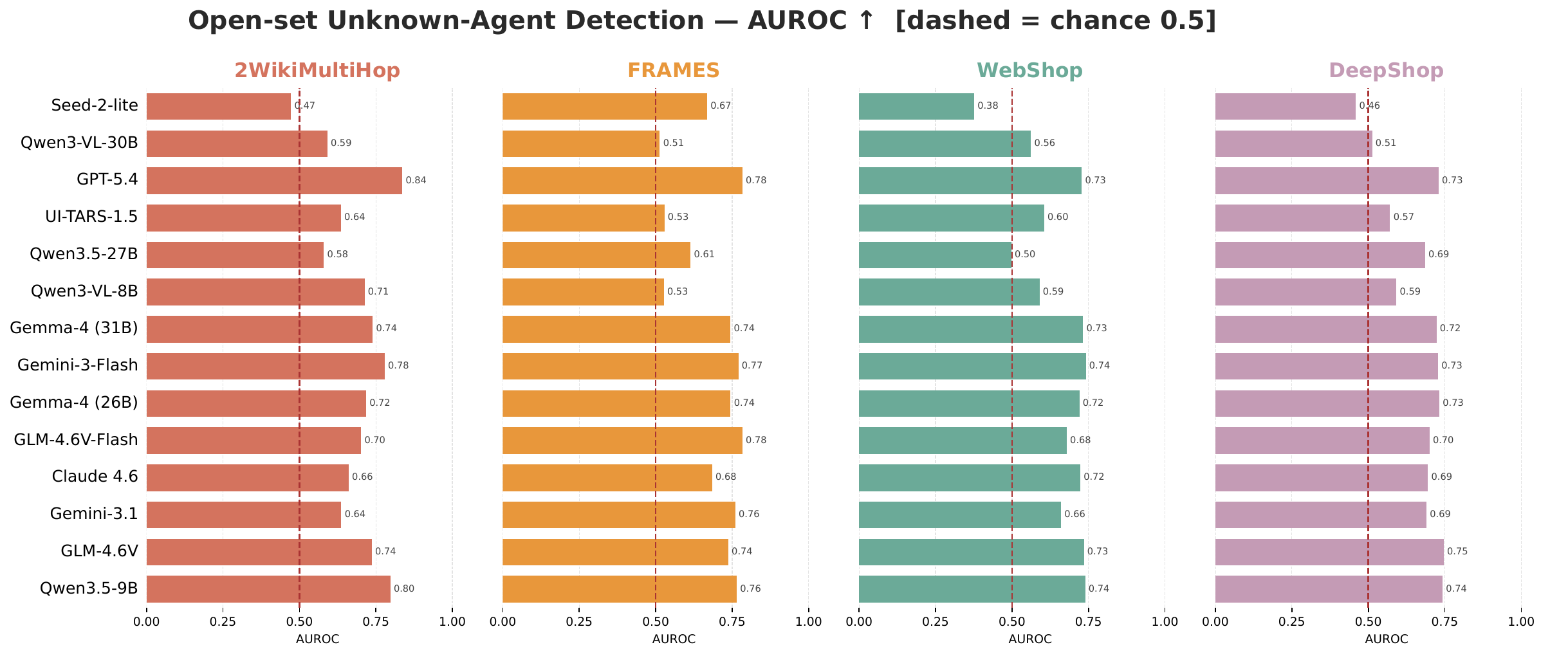}
    \caption{\textbf{Unknown agents are detectable above chance, but identification remains an open challenge.} Detection is consistently above chance across all four datasets, with the majority of agents exceeding AUROC 0.60 with our XGBoost classifier. A notable exception is Seed-2-lite, which scores below chance on 2WikiMultiHopQA, WebShop, and DeepShop despite achieving the highest closed-set identifiability. Nevertheless, robust open-set identification remains an open challenge.}
    \label{fig:main_open_bars}
\end{figure}

\subsection{Agents are identifiable from traces of their actions}

\paragraph{Known LLMs are broadly fingerprintable from their actions.} In the closed-set setting, agents are highly identifiable from action traces. Across all four benchmarks in Figure~\ref{fig:main_closed_bars}, our XGBoost classifier recovers the source model at roughly $10\times$ random chance, with per-agent F1 exceeding 70\% for the majority of models on every dataset. Top performers such as Seed-2-lite (96.1\% on 2WikiMultiHopQA) and UI-TARS-1.5 (92.1\% on WebShop) are near-perfectly identifiable, suggesting their actions are highly consistent and distinct across episodes. Performance remains high even on the weakest pair (45.4\% for GLM-4.6V-Flash on WebShop), well above the $\sim$7\% random baseline for 14 classes. This extends to family-level attribution: grouping agents by model family preserves strong identifiability without version-specific labels (Table ~\ref{tab:predicting_family}). We explore generalisation across tasks and sites in Appendix~\ref{app:site_transfer}, finding that single-task transfer is weak but pooling traces from multiple tasks on the same site recovers strong attribution.

\paragraph{Open-set fingerprinting is agent-specific and orthogonal to closed-set performance.} Detection of unknown models is consistently above chance across all four datasets and most agents, with the majority exceeding AUROC$\approx$ 0.60. However, agents that are easiest to classify when their identities are known (closed-set) are not easy to classify in an open-set setting. Most strikingly, Seed-2-lite (the best-identified agent in the closed-set setting) scores below chance on three of four datasets (AUROC 0.47 on
2WikiMultiHopQA, 0.38 on WebShop, 0.46 on DeepShop), while GPT-5.4
achieves the highest open-set AUROC overall (0.84 on 2WikiMultiHopQA)
despite ranking third in closed-set F1.

This dissociation suggests that closed-set identifiability reflects how distinct an agent is within a known distribution, whilst open-set identifiability punishes models whose behaviour isn't uniquely distinct from that of known agents.

In general, this demonstrates that open-set detection is useful even when exact attribution is impossible: for a website host, recognising that a visiting trace belongs to no currently enrolled model is sufficient to trigger offline collection and later enrollment into the fingerprint database.

\section{Analysis}\label{sec:analysis}

\subsection{Timing dominates the fingerprint, but actions are more robust to perturbation}


\begin{figure}[t]
    \centering
    \includegraphics[width=\linewidth]{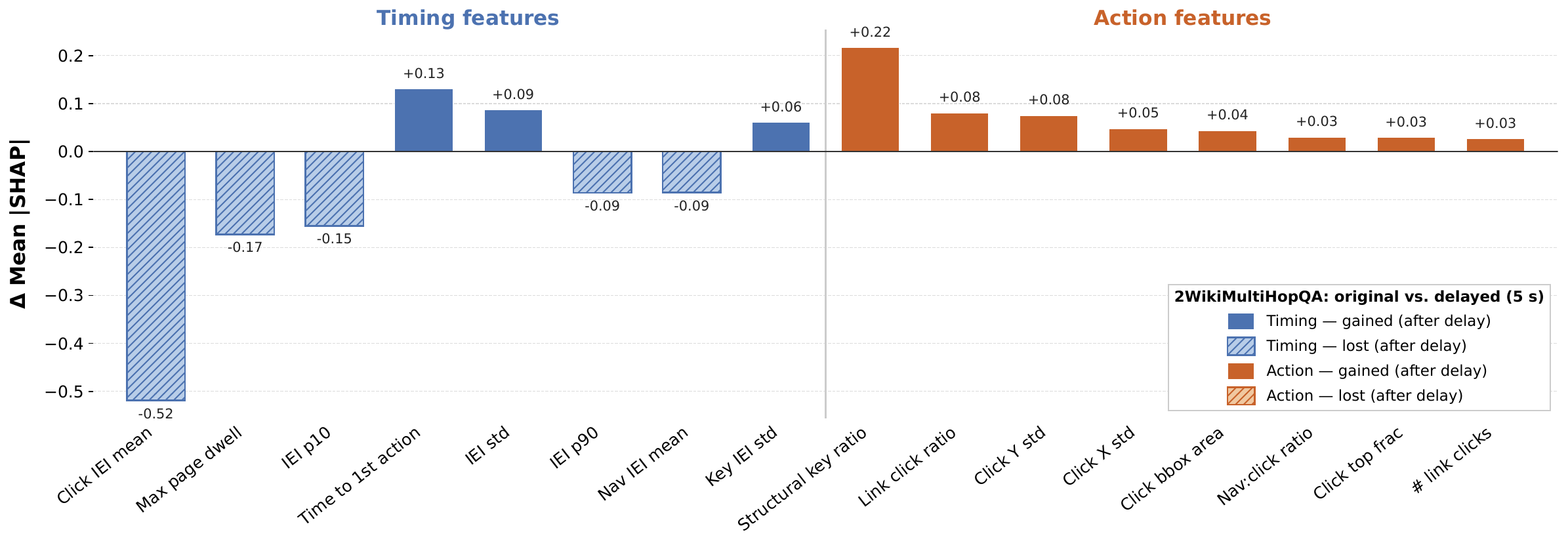}
    \caption{\textbf{Timing features are important but sensitive to perturbation, while action features are robust.} Each bar shows the change in 
mean \textbar SHAP\textbar\ for a feature when XGBoost is retrained on 
5-second-delayed traces. Under normal conditions, timing features based on IEI statistics and time to first action dominate agent identification, but after training on delayed traces, our classifier relies on action-centred features like structural key ratio and click position to make predictions.}
    \label{fig:feature_importance_shift}
\end{figure}

\paragraph{Timing is the primary signal.} We compute mean absolute SHAP values for the XGBoost classifier on 2WikiMultiHopQA before and after retraining on delayed traces in Figure \ref{fig:feature_importance_shift}. Without delays, our top features are overwhelmingly timing-based: IEI standard deviation, mean click IEI, and time to first action all receive substantially larger attributions than structural features like key ratio. Agents are distinguishable not primarily by what actions they take, but by their tempo: how long they pause before acting, how variable that pause is, and whether different action types carry their own characteristic delays. Full feature SHAP results are presented in Appendix \ref{app:feature_importance}. Aside from these features, we show that classifier performance isn't tied to overall agent capability in Appendix \ref{sec:capability-identifiability}.

\begin{figure}[t]
\centering
\includegraphics[width=\linewidth]{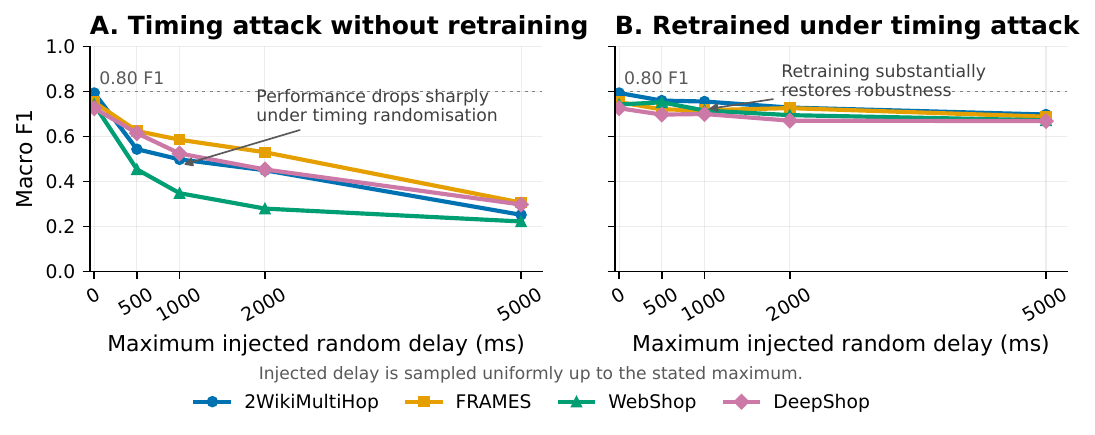}
\caption{\textbf{Simple timing delays weaken unadapted classifiers but not adaptive ones.} \textbf{(Left)} Performance of an unadapted classifier, trained on clean traces and evaluated under increasing injected delays. \textbf{(Right)} Performance of a delay-adapted classifier, retrained on delayed traces and evaluated under the same perturbation. While delay injection substantially degrades the unadapted classifier, performance largely recovers after retraining.}
\label{fig:timing_delay}
\end{figure}

\paragraph{Actions carry the fingerprint when timing is disrupted.} If classifiers rely on temporal signatures, adding random delays should be enough to break them. We test this by injecting a uniformly sampled random delay between agent actions at test time and evaluating XGBoost under increasing delay budgets in Figure \ref{fig:timing_delay}. Without retraining, macro F1 drops sharply as injected delay grows, confirming that clean-trace classifiers are sensitive to disrupted timing rhythms. However, retraining on delayed traces largely recovers performance across all four datasets. The classifier shifts weight onto features that survive delay injection: residual timing variability, click-coordinate dispersion, structural key ratio, and link-click ratio. These are features grounded in \textbf{what} agents do and, not merely \textbf{when}.

\textbf{Prompt normalisation.} We also test whether prescribing a fixed browsing procedure in the task prompt suppresses model-specific behaviour (Appendix~\ref{app:prompt-normalisation}). On WebShop, prompt normalisation lowers Macro-F1 of a classifier trained on the original prompts from 0.917 to 0.443. Retraining on normalised traces restores it to 0.955, and a classifier trained on both prompt regimes scores 0.889 and 0.910 on the original and normalised prompts respectively. Like delay injection, it hurts only a non-adaptive attacker, and it relies on models following the instruction.

\subsection{Agent fingerprinting is efficient at both training and test time}
\begin{figure}[h!]
    \centering
    \includegraphics[width=\linewidth]{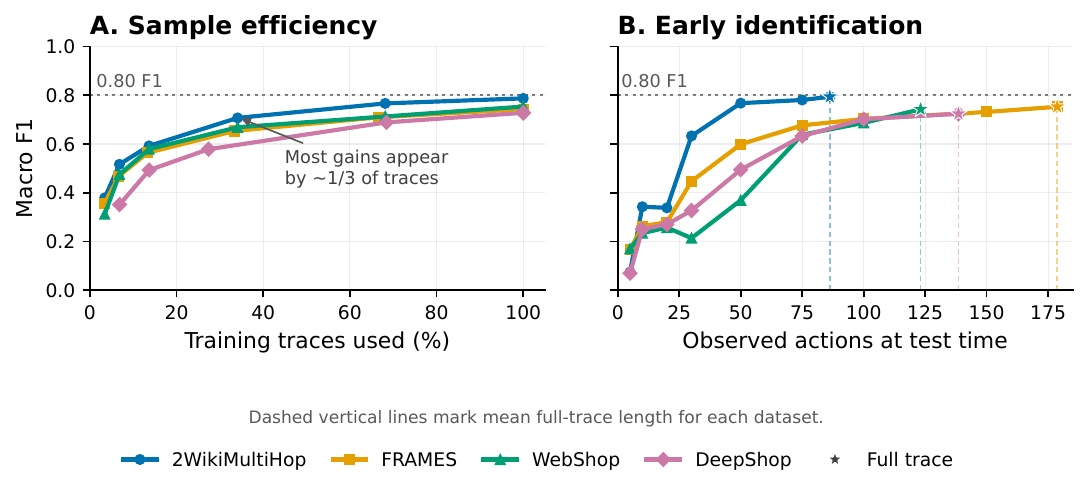}
  \caption{\textbf{Agent fingerprints emerge early and can be learned from short traces.} \textbf{(Left)} Training efficiency, measured by training classifiers using only the first $k$ events from each training trace and evaluating them on full held-out traces. \textbf{(Right)} Identification speed, measured by training classifiers on full traces and evaluating them using only the first $k$ events observed at test time. Dashed vertical 
    lines mark the mean full-trace length for each dataset.}
    \label{fig:wiki_early_graph}
\end{figure}

\paragraph{Strong classifiers can be trained from a fraction of the traces.}
By varying the proportion of training data used to fit our XGBoost classifier, we find 
that fewer than one third of traces are sufficient to approach peak classification 
performance across all four datasets (Figure~\ref{fig:wiki_early_graph}\textbf{A}). Gains 
diminish rapidly beyond this point, indicating that the behavioural signatures 
underpinning agent identity are both consistent and learnable from modest supervision.

\paragraph{Agent Identification improves steadily with observed events.}
Using our XGBoost classifier trained on full traces, we systematically reduce the number of events observed at test time to assess how quickly identity can be recovered mid-trajectory. As shown in Figure~\ref{fig:wiki_early_graph}\textbf{B}, Macro-F1 improves steadily with the number of observed events rather than saturating early. Averaged over the four datasets, it is 0.27 after 10 events and 0.56 after 50, and reaches 0.72 after 100 events, close to the 0.75 achieved on complete traces (Appendix~\ref{app:early-id}). Identification is therefore possible mid-session, with accuracy improving as the agent continues to act.

\section{Implications}\label{sec:implications}

\subsection{Attack Surface: Agent-Targeted Exploits}

Knowing the identity of a target makes an adversary's job easier, reducing the search space from all possible attacks to those known to be effective against the target \citep{hutchins_intelligence-driven_2011}. A natural application is the class of threats known as AI Agent Traps: adversarial page content designed to manipulate, deceive, or exploit visiting agents \citep{franklin_ai_2026}. Passive fingerprinting lets each trap be conditioned on model identity, converting generic exploits into targeted ones across three attack surfaces we characterise below

\paragraph{Model-specific prompt injection.}
LLMs exhibit systematically different susceptibilities to prompt injection 
\citep{aichberger_attacking_2025, pasquini_llmmap_2025}. Without identity information, 
an adversary must either deploy generic injections that may be ineffective, or mount a 
costly black-box search over model-specific failure modes, both of which are detectable 
and expensive. Action fingerprinting collapses this search \textit{passively}: once the model is 
identified, the adversary can select directly from a known corpus of model-specific 
jailbreaks \citep{shen_anything_2024, zou_universal_2023} or initialise a targeted 
optimisation with a substantially reduced search space \citep{aichberger_attacking_2025, 
pasquini_llmmap_2025, seip_preference_2026}. Because fingerprinting is passive, a visiting agent has no direct signal that it is being identified. 
Furthermore, since the injection payload can be conditioned on the inferred identity 
after fingerprinting completes, it is invisible to any visitor whose model does not 
match the target, making such attacks difficult to detect through conventional crawling 
or auditing.

\paragraph{Adversarial cost inflation.}
Sponge attacks aim to exhaust the computational budget of neural networks by inducing verbose or high-entropy reasoning \citep{shumailov_sponge_2021}, and OverThink \citep{kumar2025overthink} shows that decoy problems planted in web content can inflate a reasoning model’s token use, charging the cost to whoever runs the model. Agent identifiability introduces a targeted variant: a site operator who knows which model is visiting can serve that model pages calibrated to maximise its token consumption, through navigational ambiguity, redundant content, or structures known to trigger extended reasoning, while serving normal content to all other visitors. This constitutes a denial-of-service vector targeting the \textit{user's} inference budget rather than the host's compute, a threat class that to our knowledge has not been previously formalised. It is particularly consequential for high-cost frontier models such as Claude Opus 4.6 or GPT-5.4, where modest per-query cost increases compound significantly at scale.

\paragraph{Agent-specific access control.}
Beyond active exploitation, identity signals let site operators route content by model identity. We anticipate two near-term variants. \textit{Blacklisting} denies service to specific models for cost, legal, or competitive reasons without blocking agents broadly. \textit{Whitelisting}, the more adversarial inverse, serves model-conditioned content that misleads specific agents while appearing benign to others. This is especially dangerous because poisoned content may be invisible to auditors using a different model. Neither variant requires a fixed classifier: a site operator can \textit{continuously retrain} on organic or simulated traces, extending coverage to new models without external data. Appendix~\ref{app:site_transfer} supports this assumption: although single-task transfer is weak, pooling multiple tasks from the same website recovers strong attribution.



\section{Limitations}

\paragraph{Harness and setting.} The fingerprint is conditioned on the harness: classifiers trained on one harness do not transfer to the other, and pooled training is required (Appendix~\ref{app:harness-transfer}). We tested two harnesses, MidScene.js and Browser Use, on two websites and four tasks. Our threat model assumes that agent detection has already taken place, so we did not mix agent traces with human or other bot traffic.

\paragraph{Inference infrastructure.} In Appendix~\ref{app:engine-transfer} we varied the inference engine for three open-weight models and found that classifier performance remains high. We did not vary GPU architecture or the location of the serving nodes, either of which may change inter-event timings and hence the timing features on which our classifiers rely.

\section{Conclusion}


We show that UI actions from agent traces can fingerprint the underlying model of an LLM-based web agent with up to 96\% F1 across 14 frontier models. The signal is typically recoverable from few observed events, does not rely on network-layer visibility, and is passive. The practical consequence is a shift in the threat landscape for deployed agents. Target identification is the first step towards exploitation, and we show that this move can be executed passively during ordinary navigation by any party that controls a page visited by the agent. Every page visit thus becomes a potential attribution event on which model-specific injections, content poisoning, or budget exhaustion can be conditioned.

The operative question for the next generation of agent-aware web infrastructure is therefore no longer whether a client is human or automated, but which model is producing the behaviour. Within-agent attribution, rather than bot detection, is the axis on which both defences and cooperative protocols should be designed. We release our trace corpus and evaluation harness to make that axis measurable.

Future work should evaluate obfuscation against adaptive adversaries. In Appendix ~\ref{app:prompt-normalisation}, we show that system prompt normalisation sharply reduces classifier performance, but fails against classifiers trained on traces from agents with multiple prompt formats.

Overall, as LLM-based agents become a standard client of the web, their on-page behaviour is an identifying signal that warrants the same engineering scrutiny as the models themselves. Attackers and operators now have access to the same primitive, and the design of agent-aware infrastructure will be shaped by how each chooses to use it.
Whether it is used to exploit agents or to serve them better, the primitive is now available to both.

\section*{Acknowledgements}
LW and JM were supported by a Rhodes Scholarship. SM was supported by the AI Safety Fund and Microsoft Ltd. CR carried out this work under the ERC Horizon project DVPS.

\bibliographystyle{unsrtnat}
\bibliography{refs}

\newpage
\section*{Appendix}
\appendix
\section{Additional Experimental Details}
\subsection{LLMs Used}\label{app:sec:model_details}
\begin{table}[h!]
\centering
\caption{Models evaluated in this work.
  $^\ddagger$UI-specialist model fine-tuned for GUI agent tasks.
  Active parameter counts shown for mixture-of-experts (MoE) models.
}
\label{app:tab:models_used}
\small
\begin{tabular}{llccc}
\toprule
\textbf{Model} & \textbf{Type} & \textbf{Params} & \textbf{Active} & \textbf{Access} \\
\midrule
\multicolumn{5}{l}{\textit{Open-source --- locally hosted}} \\
\midrule
Qwen3-VL-8B \citep{team_qwen3_2025}         & General  & 8B   & 8B  & Local \\
Qwen3-VL-30B-A3B \citep{team_qwen3_2025}    & General  & 30B  & 3B  & Local \\
Qwen3.5-9B \citep{qwen_team_qwen35_2026}    & General  & 9B   & 9B  & Local \\
Qwen3.5-27B \citep{qwen_team_qwen35_2026}   & General  & 27B  & 27B & Local \\
GLM-4.6V \citep{team_glm-45v_2025}          & General  & 106B  & --- & Local \\
GLM-4.6V-Flash \citep{team_glm-45v_2025}    & General  & 9B  & --- & Local \\
UI-TARS-1.5-7B$^\ddagger$ \citep{qin_ui-tars_2025} & UI-specialist & 7B & 7B & Local \\
Gemma-4-26B-A4B \citep{gemma_team_gemma_2026} & General & 26B & 4B  & Local \\
Gemma-4-31B \citep{gemma_team_gemma_2026}   & General  & 31B  & 31B & Local \\
\midrule
\multicolumn{5}{l}{\textit{Proprietary --- API}} \\
\midrule
GPT-5.4 \citep{openai_introducing_2026}         & General & ---  & --- & API \\
Gemini-3.1 \citep{google_gemini_nodate}         & General & ---  & --- & API \\
Gemini-3-Flash \citep{google_gemini_nodate}     & General & ---  & --- & API \\
Claude Opus 4.6 \citep{anthropic_claude_nodate} & General & ---  & --- & API \\
Seed-2.0-Lite \citep{bytedance_seed_team_seed20_2026} & General & --- & --- & API \\
\bottomrule
\end{tabular}
\end{table}

\subsection{Prompt Templates for Datasets}
\begin{tcolorbox}[
  colback=purple!10,
  colframe=gray!50,
  boxrule=0.5pt,
  arc=2pt,
  boxsep=5pt,
  left=8pt,
  right=8pt,
  top=6pt,
  bottom=6pt
]
\small
\textbf{Task: Wikipedia Question Answering}

\vspace{2pt}
You are browsing Wikipedia to answer a question.

\vspace{4pt}
\textbf{Rules:}
\begin{itemize}
  \setlength{\itemsep}{2pt}
  \setlength{\parskip}{0pt}
  \setlength{\parsep}{0pt}
  \item Restrict browsing to \texttt{https://www.wikipedia.org}.
  \item Use the browser to gather evidence before answering.
  \item Output the final answer as \texttt{<answer>...</answer>}.
\end{itemize}

\vspace{4pt}
\textbf{Question:}

\vspace{2pt}
\texttt{\{question\}}
\end{tcolorbox}

\begin{tcolorbox}[
  colback=orange!10,
  colframe=gray!50,
  boxrule=0.5pt,
  arc=2pt,
  boxsep=5pt,
  left=8pt,
  right=8pt,
  top=6pt,
  bottom=6pt
]
\small
\textbf{Task: Amazon Product Selection}

\vspace{2pt}
You are a shopping assistant tasked with finding the product that best matches the description.

\vspace{4pt}
\textbf{Instructions:}
\begin{itemize}
  \setlength{\itemsep}{2pt}
  \setlength{\parskip}{0pt}
  \setlength{\parsep}{0pt}
  \item Browse Amazon freely (search, filter, and view product pages).
  \item Add promising items to your cart and refine your selection.
  \item Remove items if better alternatives are found.
  \item Do not log in or proceed to checkout.
\end{itemize}

\vspace{4pt}
At completion, your cart should contain only the product(s) that best match the description.

\vspace{4pt}
\textbf{Description:}

\vspace{2pt}
\texttt{\{description\}}
\end{tcolorbox}








\subsection{Classifier Hyperparameters}\label{app:classifier_hyperparams}
We train five classifier families on the behavioural feature set described in Section~\ref{app:behavioral_features}: a Random Forest, XGBoost, two Logistic Regression variants (L2 and L1), and a hybrid LSTM. All classifiers are selected by validation accuracy via cross-validated hyperparameter search; final models are refit on the training split only.

\paragraph{Random Forest.}
Exhaustive grid search (3-fold CV) over the space in Table~\ref{tab:rf_grid}. Features are used without scaling.

\begin{table}[h]
\centering
\small
\begin{tabular}{ll}
\toprule
\textbf{Hyperparameter} & \textbf{Search space} \\
\midrule
\texttt{n\_estimators}      & 200, 400 \\
\texttt{max\_depth}         & None (unlimited), 15, 30 \\
\texttt{max\_features}      & \texttt{sqrt}, \texttt{log2}, 0.4 \\
\texttt{min\_samples\_split} & 2, 5 \\
\bottomrule
\end{tabular}
\caption{Random Forest hyperparameter grid (exhaustive, $2{\times}3{\times}3{\times}2 = 36$ configurations).}
\label{tab:rf_grid}
\end{table}

\paragraph{XGBoost.}
Randomised search (40 iterations, 3-fold CV, \texttt{hist} tree method) over the space in Table~\ref{tab:xgb_grid}. Features are used without scaling.

\begin{table}[h]
\centering
\small
\begin{tabular}{ll}
\toprule
\textbf{Hyperparameter} & \textbf{Search space} \\
\midrule
\texttt{n\_estimators}     & 100, 200, 300, 400, 500 \\
\texttt{learning\_rate}    & 0.01, 0.05, 0.1, 0.2, 0.3 \\
\texttt{max\_depth}        & 3, 4, 5, 6, 7, 8 \\
\texttt{subsample}         & 0.6, 0.7, 0.8, 0.9, 1.0 \\
\texttt{colsample\_bytree} & 0.5, 0.6, 0.7, 0.8, 1.0 \\
\texttt{reg\_alpha}        & 0, 0.01, 0.1, 1.0 \\
\texttt{reg\_lambda}       & 0.5, 1.0, 2.0, 5.0 \\
\bottomrule
\end{tabular}
\caption{XGBoost hyperparameter search space (randomised, 40 draws).}
\label{tab:xgb_grid}
\end{table}

\paragraph{Logistic Regression.}
We train two variants with 3-fold grid search over $C \in \{0.01, 0.1, 1.0, 10.0\}$, with a maximum of 5{,}000 iterations. Features are $z$-score normalised (StandardScaler fit on training data only). LR-L2 uses the \texttt{lbfgs} solver; LR-Lasso uses the \texttt{saga} solver with full L1 regularisation.

\paragraph{LSTM.}
The AgentLSTM encodes the raw browser event sequence. Each event is represented as a token embedding (dimension 16) concatenated with five continuous scalars: log inter-event gap, log absolute session timestamp, spatial position (normalised click $x/y$ or scroll depth), and log running per-type inter-event interval. The token vocabulary covers eight event types (click, keydown, scroll, navigate, beforeunload, focus, plus \texttt{<pad>} and \texttt{<unk>}). The LSTM final hidden state is concatenated with the pre-computed aggregate feature vector before the classification head.

Fixed training hyperparameters are listed in Table~\ref{tab:lstm_fixed}; the grid searched over hidden dimension and dropout is given in Table~\ref{tab:lstm_grid}. Best configuration is selected by validation accuracy; the final model is refit from scratch on the training split for 50 epochs.

\begin{table}[h]
\centering
\small
\begin{tabular}{ll}
\toprule
\textbf{Hyperparameter} & \textbf{Value} \\
\midrule
Embedding dimension      & 16 \\
Number of LSTM layers    & 2 \\
Optimizer                & Adam \\
Learning rate            & $1 \times 10^{-3}$ \\
Weight decay             & $1 \times 10^{-4}$ \\
Batch size               & 16 \\
Epochs                   & 50 \\
Loss                     & Cross-entropy \\
\bottomrule
\end{tabular}
\caption{Fixed LSTM training hyperparameters.}
\label{tab:lstm_fixed}
\end{table}

\begin{table}[h]
\centering
\small
\begin{tabular}{ll}
\toprule
\textbf{Hyperparameter} & \textbf{Search space} \\
\midrule
Hidden dimension & 64, 128, 256, 512 \\
Dropout          & 0.2, 0.4 \\
\bottomrule
\end{tabular}
\caption{LSTM grid search space ($4{\times}2 = 8$ configurations, selected by validation accuracy).}
\label{tab:lstm_grid}
\end{table}

\subsection{Harness Configuration}\label{app:harness_config}
\paragraph{Overview.}
Each episode is executed by a Python orchestrator (\texttt{orchestrator.py}) that dispatches agent runs as isolated Apptainer container invocations. The container (\texttt{agent.sif}) packages a headless Chromium browser, the Playwright automation framework, and the MidScene \texttt{PlaywrightAgent} action loop. This design ensures that no browser state, cookies, or cached content persists across episodes or across agents.

\paragraph{Episode execution.}
For each (agent, question) pair, the orchestrator invokes the container with the agent's API credentials, the task prompt, and a unique episode ID. Inside the container, \texttt{agent\_runner.ts} navigates Chromium to the task start URL and runs \texttt{agent.aiAct(prompt)} until the task is complete or the per-episode timeout is reached. The browser viewport is fixed at $1280 \times 768$ pixels across all agents and tasks. The MidScene replanning cycle limit is set to 40 across all agents, capping the maximum number of action--observation loops per episode.

\paragraph{Event collection.}
Browser interaction events are captured via two complementary paths to ensure completeness:
\begin{itemize}
  \item \textbf{Primary (push bridge):} A \texttt{page\_tracer.js} init script is injected into every page context. It patches the DOM event listeners and calls \texttt{window.\_\_pushTraceEvent} for each recorded interaction (click, keydown, scroll, navigate, beforeunload, focus). Playwright routes these calls over the Chrome DevTools Protocol (CDP) to a host-side handler, which timestamps each event relative to the episode wall-clock start and appends it to the episode buffer. No polling is required.
  \item \textbf{Secondary (backstop harvest):} A single \texttt{harvest()} call at episode end reads any events remaining in the in-page \texttt{\_\_agentTrace.events} array that did not complete their CDP bridge call before page teardown (most commonly \texttt{beforeunload} events mid-navigation).
\end{itemize}
HTTP-level page navigations — invisible to the in-page tracer — are captured separately via Playwright's \texttt{framenavigated} listener and appended as \texttt{navigate} events with a \texttt{trigger: "http"} field.

\paragraph{Output format.}
Each completed episode is written as a JSON file at:
\begin{center}
\texttt{traces/\{agent\_id\}/\{dataset\_name\}/\{timestamp\}/\{episode\_id\}.json}
\end{center}
The file contains episode metadata (agent ID, model name, task type, timestamp), the agent's answer and optional verification result, the MidScene action log, and the full DOM event trace including all timestamped browser interactions.

\paragraph{Run configuration.}\label{app:trace_collection_run_configuration}
Experiments are specified via YAML configuration files that define the agent set, dataset slices, and run parameters. Table~\ref{tab:harness_config} summarises the fixed run parameters used across all experiments reported in this paper. Collecting traces for all datasets takes approximately three days. The cost of generating text for API accessed models totals USD 2890. The API's used are provided by Anthropic, OpenAI and OpenRouter.

\begin{table}[h!]
\centering
\small
\begin{tabular}{ll}
\toprule
\textbf{Parameter} & \textbf{Value} \\
\midrule
Browser                        & Chromium (headless) \\
Viewport                       & $1280 \times 768$ px \\
Agent framework                & MidScene \texttt{PlaywrightAgent} \\
Replanning cycle limit         & 40 \\
Episodes per (agent, question) & 1 \\
Per-episode timeout            & 300\,s \\
Parallel workers               & 5 \\
Container runtime              & Apptainer \\
\bottomrule
\end{tabular}
\caption{Fixed harness parameters used in all experiments.}
\label{tab:harness_config}
\end{table}

\subsection{Behavioural Features Collected}
\label{app:behavioral_features}

For each browsing episode, we extract a fixed-dimensional feature vector from
the client-side event trace. Let an episode consist of an ordered sequence of
DOM events
\[
E = \{e_1, e_2, \dots, e_T\},
\]
where each event has a type (e.g., click, scroll, navigate), a timestamp, and
optionally additional fields such as URL, scroll percentage, or screen
coordinates. We also use the corresponding Midscene action log, which records
the high-level actions issued by the agent through the browser harness.


\paragraph{Event subsets.}
From the raw event sequence, we define the following subsets:
\[
E_{\text{click}},\;
E_{\text{scroll}},\;
E_{\text{nav}},\;
E_{\text{keydown}},\;
E_{\text{beforeunload}},\;
E_{\text{focus}},
\]
corresponding respectively to click, scroll, navigation, keydown,
\texttt{beforeunload}, and focus events. Let $M$ denote the Midscene action
log for the same episode.

We define the page count $P$ as the number of distinct pages visited in the
episode, using the recorded page count when available and otherwise falling
back to the number of unique URLs observed in the DOM trace.

\paragraph{Temporal dynamics.}
To characterise interaction rhythm, we extract the sequence of event timestamps
\[
t_1, t_2, \dots, t_T,
\]
and compute inter-event intervals
\[
\Delta_i = t_{i+1} - t_i \quad \text{for } i=1,\dots,T-1.
\]

\paragraph{Behavioral Feature Set}\label{app:behavioral_features}

We extract 41 scalar features from each episode's browser event trace (DOM event log). Features are grouped into seven categories below. All features are computed purely from client-side browser events (click, keydown, scroll, navigate, beforeunload, focus) and require no access to model internals or outputs.

\begin{longtable}{lp{7.5cm}}
\toprule
\textbf{Feature} & \textbf{Description} \\
\midrule
\multicolumn{2}{l}{\textit{Event volume}} \\
\midrule
\texttt{n\_clicks}           & Total number of click events \\
\texttt{n\_scrolls}          & Total number of scroll events \\
\texttt{n\_navigations}      & Total number of page navigation events \\
\texttt{n\_keydowns}         & Total number of keydown events \\
\texttt{n\_focus}            & Total number of input/textarea focus events \\
\texttt{n\_events\_total}    & Total events across all types \\
\texttt{page\_count}         & Number of distinct pages visited \\
\texttt{n\_unique\_domains}  & Number of unique hostnames visited \\
\midrule
\multicolumn{2}{l}{\textit{Global timing}} \\
\midrule
\texttt{total\_duration\_s}   & Wall-clock episode duration (seconds) \\
\texttt{t\_first\_action\_ms} & Time from episode start to first event (ms) \\
\texttt{mean\_iei\_ms}        & Mean inter-event interval across all events (ms) \\
\texttt{std\_iei\_ms}         & Standard deviation of inter-event intervals (ms) \\
\texttt{median\_iei\_ms}      & Median inter-event interval (ms) \\
\texttt{p10\_iei\_ms}         & 10th percentile inter-event interval (ms) \\
\texttt{p90\_iei\_ms}         & 90th percentile inter-event interval (ms) \\
\texttt{iei\_trend}           & Ratio of mean IEI in the second half of the episode to the first half; values $>1$ indicate the agent slows down as context grows \\
\midrule
\multicolumn{2}{l}{\textit{Per-type planning latency}} \\
\midrule
\texttt{mean\_click\_iei\_ms} & Mean inter-click interval (ms) \\
\texttt{std\_click\_iei\_ms}  & Std.\ of inter-click intervals (ms) \\
\texttt{mean\_nav\_iei\_ms}   & Mean inter-navigation interval, approximating page dwell time (ms) \\
\texttt{std\_nav\_iei\_ms}    & Std.\ of inter-navigation intervals (ms) \\
\texttt{max\_page\_dwell\_ms} & Maximum single-page dwell time (ms) \\
\texttt{mean\_key\_iei\_ms}   & Mean inter-keydown interval, approximating API keystroke latency (ms) \\
\texttt{std\_key\_iei\_ms}    & Std.\ of inter-keydown intervals (ms) \\
\midrule
\multicolumn{2}{l}{\textit{Scroll behavior}} \\
\midrule
\texttt{max\_scroll\_pct}     & Maximum scroll depth reached, as a percentage of page height \\
\texttt{mean\_scroll\_pct}    & Mean scroll depth across all scroll events \\
\texttt{n\_deep\_scrolls}     & Number of scroll events reaching $>$60\% page depth \\
\texttt{scroll\_reversals}    & Number of direction reversals in the scroll depth sequence \\
\midrule
\multicolumn{2}{l}{\textit{Click spatial distribution}} \\
\midrule
\texttt{click\_x\_std}           & Standard deviation of click $x$-coordinates (pixels) \\
\texttt{click\_y\_std}           & Standard deviation of click $y$-coordinates (pixels) \\
\texttt{click\_bbox\_area\_frac} & Bounding-box area of all click positions as a fraction of the $1280{\times}768$ viewport \\
\texttt{click\_top\_frac}        & Fraction of clicks in the top quarter of the viewport ($y < 192$px), capturing navbar/search-bar interaction \\
\texttt{n\_link\_clicks}         & Number of clicks on anchor elements (href present) \\
\texttt{link\_click\_ratio}      & \texttt{n\_link\_clicks} / \texttt{n\_clicks} \\
\midrule
\multicolumn{2}{l}{\textit{Navigation strategy}} \\
\midrule
\texttt{popstate\_ratio}         & Fraction of navigations triggered by \texttt{popstate} (history back) \\
\texttt{scroll\_to\_click\_ratio}& \texttt{n\_scrolls} / \texttt{n\_clicks} \\
\texttt{actions\_per\_page}      & \texttt{n\_events\_total} / \texttt{page\_count} \\
\texttt{nav\_to\_click\_ratio}   & \texttt{n\_navigations} / \texttt{n\_clicks} \\
\texttt{keydowns\_per\_page}     & \texttt{n\_keydowns} / \texttt{page\_count} \\
\texttt{focus\_per\_page}        & \texttt{n\_focus} / \texttt{page\_count} \\
\texttt{structural\_key\_ratio}  & Fraction of keydowns that are structural keys (Enter, Arrow*, Tab, Escape, Backspace, Delete) vs.\ printable characters \\
\midrule
\multicolumn{2}{l}{\textit{Exit behavior}} \\
\midrule
\texttt{mean\_exit\_scroll\_pct} & Mean scroll depth at \texttt{beforeunload} events, reflecting how far the agent had read before leaving each page \\
\bottomrule
\caption{The 41 behavioral features extracted from each episode trace}
\label{tab:features}
\end{longtable}

\section{Supplementary Results}
\small
\begin{longtable}{l l c c c c}

\caption{Agent identification  F1 (\%) across datasets and classifiers. Best F1 per model group in \textbf{bold}. Macro f1 for each dataset and classifier is at the bottom of the table.}
\label{app:tab:main} \\

\toprule
\textbf{Model} & \textbf{Clf.} &
  \makecell{\textbf{2Wiki} \\ \textit{(in-dom.)}} &
  \makecell{\textbf{FRAMES} \\ \textit{(in-dom.)}} &
  \makecell{\textbf{Webshop} \\ \textit{(in-dom.)}} &
  \makecell{\textbf{DeepShop} \\ \textit{(in-dom.)}} \\
\midrule
\endfirsthead

\multicolumn{6}{c}{\tablename~\thetable{} -- \textit{continued from previous page}} \\[4pt]
\toprule
\textbf{Model} & \textbf{Clf.} &
  \makecell{\textbf{2Wiki} \\ \textit{(in-dom.)}} &
  \makecell{\textbf{FRAMES} \\ \textit{(in-dom.)}} &
  \makecell{\textbf{Webshop} \\ \textit{(in-dom.)}} &
  \makecell{\textbf{DeepShop} \\ \textit{(in-dom.)}} \\
\midrule
\endhead

\midrule
\multicolumn{6}{r}{\textit{Continued on next page}} \\
\endfoot

\bottomrule
\endlastfoot

\rowcolor{headergreen}
\multicolumn{6}{l}{\textbf{\textit{Proprietary Models}}} \\
\midrule
GPT-5.4              & RF   & 85.71 & 72.19 & 68.09 & 66.67 \\
                     & XGB  & \textbf{91.50} & \textbf{78.48} & \textbf{68.78} & \textbf{68.29} \\
                     & LSTM & 76.19 & 73.85 & 65.90 & 46.15 \\
                     & Lasso & 78.53 & 66.23 & 61.29 & 58.67 \\
                     & LR   & 77.38 & 64.47 & 61.96 & 61.11 \\
\midrule
Claude Opus 4.6      & RF   & 67.90 & 58.54 & \textbf{67.13} & 56.34 \\
                     & XGB  & \textbf{70.51} & \textbf{68.71} & 66.18 & \textbf{75.00} \\
                     & LSTM & 46.51 & 58.03 & 57.14 & 48.72 \\
                     & Lasso & 52.11 & 59.67 & 55.81 & 57.83 \\
                     & LR   & 47.89 & 60.34 & 56.49 & 60.24 \\
\midrule
Gemini-3.1-Pro       & RF   & 64.75 & 64.86 & \textbf{75.17} & 52.05 \\
                     & XGB  & \textbf{65.19} & \textbf{75.16} & 71.43 & 59.46 \\
                     & LSTM & 61.73 & 59.88 & 72.48 & \textbf{61.11} \\
                     & Lasso & 54.29 & 57.55 & 70.67 & 42.25 \\
                     & LR   & 51.47 & 58.16 & 68.00 & 43.24 \\
\midrule
Gemini-3-Flash       & RF   & 66.67 & \textbf{54.01} & 66.23 & 61.33 \\
                     & XGB  & \textbf{76.12} & 52.70 & \textbf{71.14} & \textbf{74.29} \\
                     & LSTM & 67.53 & 49.18 & 56.16 & 55.88 \\
                     & Lasso & 67.10 & 37.18 & 59.15 & 46.58 \\
                     & LR   & 67.53 & 40.26 & 59.57 & 43.24 \\
\midrule
Seed-2.0-Lite        & RF   & 93.51 & 95.36 & 93.33 & \textbf{90.24} \\
                     & XGB  & \textbf{96.05} & 96.60 & 93.42 & 87.50 \\
                     & LSTM & 87.90 & \textbf{96.69} & \textbf{96.05} & 90.14 \\
                     & Lasso & 89.47 & 87.84 & 95.17 & 86.42 \\
                     & LR   & 92.62 & 87.84 & 95.17 & 85.00 \\
\midrule
\rowcolor{headerblue}
\multicolumn{6}{l}{\textbf{\textit{Open-Source Models}}} \\
\midrule
Gemma-4-31B-it       & RF   & 80.26 & 73.68 & 73.55 & 53.16 \\
                     & XGB  & \textbf{82.12} & 82.80 & \textbf{74.17} & 57.89 \\
                     & LSTM & 73.37 & \textbf{84.77} & 67.69 & \textbf{63.41} \\
                     & Lasso & 81.38 & 79.74 & 60.92 & 55.56 \\
                     & LR   & 81.88 & 77.63 & 61.99 & 56.34 \\
\midrule
Gemma-4-26B          & RF   & 69.92 & 48.65 & 61.64 & 30.19 \\
                     & XGB  & \textbf{72.87} & 58.11 & \textbf{72.26} & \textbf{57.58} \\
                     & LSTM & 52.17 & \textbf{64.38} & 58.76 & 38.60 \\
                     & Lasso & 56.91 & 52.70 & 58.39 & 53.12 \\
                     & LR   & 56.67 & 51.35 & 57.35 & 53.12 \\
\midrule
GLM-4.6V             & RF   & 60.43 & 54.01 & 75.52 & 67.47 \\
                     & XGB  & \textbf{64.56} & \textbf{61.54} & 79.45 & \textbf{75.61} \\
                     & LSTM & 38.46 & 48.18 & \textbf{80.75} & 67.57 \\
                     & Lasso & 53.59 & 56.00 & 77.50 & 62.50 \\
                     & LR   & 51.28 & 59.38 & 75.61 & 65.82 \\
\midrule
GLM-4.6V-Flash       & RF   & 63.44 & 54.41 & \textbf{54.10} & 68.29 \\
                     & XGB  & 70.86 & 65.28 & 45.38 & \textbf{72.73} \\
                     & LSTM & \textbf{81.88} & \textbf{93.24} & 35.59 & 60.24 \\
                     & Lasso & 56.25 & 57.33 & 24.24 & 59.52 \\
                     & LR   & 57.67 & 56.38 & 27.27 & 56.47 \\
\midrule
Qwen3-VL-30B         & RF   & 92.52 & \textbf{81.82} & 63.16 & 77.78 \\
                     & XGB  & \textbf{92.72} & 80.27 & \textbf{73.53} & \textbf{83.78} \\
                     & LSTM & 88.61 & 79.75 & 57.72 & 61.11 \\
                     & Lasso & 85.16 & 70.00 & 56.58 & 70.13 \\
                     & LR   & 87.01 & 70.89 & 59.21 & 69.23 \\
\midrule
Qwen3-VL-8B          & RF   & \textbf{82.61} & \textbf{83.45} & 70.73 & \textbf{68.42} \\
                     & XGB  & \textbf{82.61} & 81.38 & \textbf{75.15} & \textbf{68.42} \\
                     & LSTM & 67.20 & 74.45 & 57.99 & 40.54 \\
                     & Lasso & 66.67 & 68.57 & 51.85 & 44.78 \\
                     & LR   & 68.09 & 70.42 & 51.09 & 41.18 \\
\midrule
Qwen3.5-27B          & RF   & 84.15 & 94.94 & 80.56 & 86.49 \\
                     & XGB  & \textbf{90.91} & 97.40 & \textbf{82.19} & \textbf{87.67} \\
                     & LSTM & 64.20 & \textbf{98.67} & 70.50 & 75.32 \\
                     & Lasso & 74.84 & 97.40 & 74.17 & 86.11 \\
                     & LR   & 76.25 & 96.77 & 73.47 & 84.93 \\
\midrule
Qwen3.5-9B           & RF   & 58.72 & 55.12 & 73.02 & 59.70 \\
                     & XGB  & \textbf{63.72} & \textbf{65.22} & \textbf{74.07} & \textbf{64.86} \\
                     & LSTM & 38.98 & 56.29 & 58.16 & 50.49 \\
                     & Lasso & 42.11 & 48.12 & 60.87 & 59.74 \\
                     & LR   & 44.64 & 48.89 & 59.57 & 60.53 \\
\midrule
UI-TARS-1.5-7B       & RF   & 89.93 & 88.20 & 90.14 & 77.78 \\
                     & XGB  & \textbf{91.28} & \textbf{90.07} & \textbf{92.09} & \textbf{82.86} \\
                     & LSTM & 89.80 & 86.71 & 77.04 & 74.29 \\
                     & Lasso & 84.00 & 70.44 & 75.18 & 71.43 \\
                     & LR   & 86.11 & 72.96 & 75.18 & 72.46 \\
\midrule
\rowcolor{headergreen}
\multicolumn{6}{l}{\textbf{\textit{Macro F1 (14 models)}}} \\
\midrule
All                  & RF   & 75.75 & 69.95 & 72.31 & 65.42 \\
                     & XGB  & \textbf{79.36} & \textbf{75.27} & \textbf{74.23} & \textbf{72.57} \\
                     & LSTM & 66.75 & 73.15 & 65.14 & 59.54 \\
                     & Lasso & 67.31 & 64.91 & 62.99 & 61.05 \\
                     & LR   & 67.61 & 65.41 & 63.00 & 60.92 \\

\end{longtable}

\begin{table}[h!]
\caption{OOD Per-Agent identification F1 (\%) across datasets and classifiers. We report the Macro F1 for each classifier at the bottom. Best F1 per model group in \textbf{bold}}
\label{app:tab:main_ood}
\centering
\renewcommand{\arraystretch}{1.15}
\setlength{\tabcolsep}{5pt}
\resizebox{\textwidth}{!}{%
\begin{tabular}{l l c c c c }
\toprule
\textbf{Model} & \textbf{Clf.} & \makecell{\textbf{2Wiki$\to$FRAMES} \\ \textit{(OOD)}} & \makecell{\textbf{FRAMES$\to$2Wiki} \\ \textit{(OOD)}} & \makecell{\textbf{DeepShop$\to$Webshop} \\ \textit{(OOD)}} & \makecell{\textbf{Webshop$\to$DeepShop} \\ \textit{(OOD)}} \\
\midrule
\rowcolor{headergreen}
\multicolumn{6}{l}{\textbf{\textit{Proprietary Models}}} \\
\midrule
GPT-5.4              & RF   & 17.58 & \textbf{51.11} & \textbf{65.53} & 59.71 \\
                     & XGB  & 17.02 & 50.97 & 60.43 & \textbf{62.05} \\
                     & LSTM & \textbf{36.30} & 39.74 & 48.05 & 60.34 \\
\midrule
Claude Opus 4.6      & RF   & 23.24 & 31.47 & \textbf{50.37} & 37.91 \\
                     & XGB  & \textbf{26.87} & 30.69 & 46.65 & 37.22 \\
                     & LSTM & 14.61 & \textbf{35.04} & 27.42 & \textbf{38.50} \\
\midrule
Gemini-3.1-Pro       & RF   & 16.62 & 32.07 & \textbf{58.89} & 40.58 \\
                     & XGB  & 21.23 & 34.66 & 57.05 & 38.77 \\
                     & LSTM & \textbf{21.73} & \textbf{37.75} & 54.11 & \textbf{52.91} \\
\midrule
Gemini-3-Flash       & RF   & 36.96 & \textbf{55.08} & 39.95 & 40.49 \\
                     & XGB  & \textbf{37.78} & 53.45 & \textbf{52.85} & \textbf{53.39} \\
                     & LSTM & 34.62 & 45.56 & 40.72 & 44.37 \\
\midrule
Seed-2.0-Lite        & RF   & 67.05 & 86.05 & 89.03 & 74.61 \\
                     & XGB  & 69.79 & \textbf{90.73} & \textbf{89.60} & \textbf{75.77} \\
                     & LSTM & \textbf{71.39} & 84.73 & 48.80 & 73.74 \\
\midrule
\rowcolor{headerblue}
\multicolumn{6}{l}{\textbf{\textit{Open-Source Models}}} \\
\midrule
Gemma-4-31B-it       & RF   & 49.01 & 42.11 & 46.89 & 44.84 \\
                     & XGB  & 43.61 & 44.34 & \textbf{48.26} & 43.75 \\
                     & LSTM & \textbf{52.37} & \textbf{51.94} & 40.68 & \textbf{52.57} \\
\midrule
Gemma-4-26B          & RF   & 35.19 & 51.39 & 21.34 & 25.00 \\
                     & XGB  & \textbf{35.77} & \textbf{55.26} & \textbf{28.51} & \textbf{35.78} \\
                     & LSTM & 35.25 & 53.58 & 15.22 & 22.83 \\
\midrule
GLM-4.6V             & RF   & 35.15 & 22.87 & 25.41 & \textbf{65.08} \\
                     & XGB  & \textbf{35.73} & \textbf{30.93} & \textbf{60.08} & 64.80 \\
                     & LSTM & 34.61 & 21.56 & 19.23 & 43.65 \\
\midrule
GLM-4.6V-Flash       & RF   & 41.16 & 48.18 & \textbf{54.58} & \textbf{23.70} \\
                     & XGB  & 44.44 & 44.22 & 42.93 & 21.80 \\
                     & LSTM & \textbf{80.13} & \textbf{80.89} & 37.26 & 18.78 \\
\midrule
Qwen3-VL-30B         & RF   & 67.11 & 75.37 & 44.16 & 55.32 \\
                     & XGB  & \textbf{67.34} & 77.01 & \textbf{49.28} & \textbf{60.50} \\
                     & LSTM & 66.53 & \textbf{77.44} & 26.73 & 50.22 \\
\midrule
Qwen3-VL-8B          & RF   & \textbf{73.74} & 59.96 & 55.16 & 60.06 \\
                     & XGB  & 71.28 & \textbf{64.73} & \textbf{57.34} & \textbf{64.90} \\
                     & LSTM & 63.91 & 54.04 & 44.37 & 57.06 \\
\midrule
Qwen3.5-27B          & RF   & 0.15 & 0.61 & \textbf{64.68} & \textbf{63.09} \\
                     & XGB  & 1.86 & 0.62 & 63.22 & 57.14 \\
                     & LSTM & \textbf{21.98} & \textbf{10.98} & 28.13 & 41.57 \\
\midrule
Qwen3.5-9B           & RF   & 30.81 & 36.59 & 40.36 & 32.84 \\
                     & XGB  & \textbf{31.25} & 36.95 & \textbf{44.09} & 39.07 \\
                     & LSTM & 30.50 & \textbf{38.81} & 30.61 & \textbf{47.55} \\
\midrule
UI-TARS-1.5-7B       & RF   & \textbf{74.72} & 84.05 & 80.29 & 80.81 \\
                     & XGB  & 71.05 & 82.12 & \textbf{81.82} & \textbf{83.15} \\
                     & LSTM & 70.06 & \textbf{85.62} & 77.44 & 79.42 \\
\midrule
\rowcolor{headergreen}
\multicolumn{6}{l}{\textbf{\textit{All Models (Macro F1)}}} \\
\midrule
All                  & RF   & 40.61 & 48.35 & 52.62 & 50.29 \\
                     & XGB  & 41.07 & 49.76 & \textbf{55.87} & \textbf{52.72} \\
                     & LSTM & \textbf{45.28} & \textbf{51.26} & 38.48 & 48.82 \\
\bottomrule
\end{tabular}%
}
\end{table}

\begin{table}[t]
\centering
\caption{\textbf{Model families are also highly identifiable from
their action traces.} There is a slight increase in overall  performance for our XGBoost classifier when predicting by model families.}
\label{tab:predicting_family}
\setlength{\tabcolsep}{5pt}
\begin{tabular}{@{}L{3.55cm}*{4}{C{2.3cm}}@{}}
\toprule
\textbf{\normalsize Family} &
\textbf{\normalsize 2Wiki} &
\textbf{\normalsize FRAMES} &
\textbf{\normalsize WebShop} &
\textbf{\normalsize DeepShop} \\
\midrule
\modelcell{Seed-2}
  & \heatcellscore{95.3}
  & \heatcellscore{95.9}
  & \heatcellscore{93.9}
  & \heatcellscore{87.2} \\
\modelcell{GPT-5}
  & \heatcellscore{90.2}
  & \heatcellscore{80.8}
  & \heatcellscore{74.4}
  & \heatcellscore{69.1} \\
\modelcell{Claude 4}
  & \heatcellscore{64.4}
  & \heatcellscore{65.8}
  & \heatcellscore{67.2}
  & \heatcellscore{73.5} \\
\modelcell{Gemini-3}
  & \heatcellscore{68.2}
  & \heatcellscore{72.5}
  & \heatcellscore{75.2}
  & \heatcellscore{60.9} \\
\modelcell{Gemini-3-Flash}
  & \heatcellscore{77.2}
  & \heatcellscore{52.4}
  & \heatcellscore{71.8}
  & \heatcellscore{68.5} \\
\modelcell{Gemma-4}
  & \heatcellscore{86.5}
  & \heatcellscore{77.6}
  & \heatcellscore{76.7}
  & \heatcellscore{69.9} \\
\modelcell{GLM-4.6V}
  & \heatcellscore{68.4}
  & \heatcellscore{75.9}
  & \heatcellscore{67.1}
  & \heatcellscore{71.0} \\
\modelcell{Qwen3-VL}
  & \heatcellscore{90.7}
  & \heatcellscore{94.7}
  & \heatcellscore{91.3}
  & \heatcellscore{87.3} \\
\modelcell{Qwen3.5}
  & \heatcellscore{75.4}
  & \heatcellscore{76.6}
  & \heatcellscore{79.6}
  & \heatcellscore{76.4} \\
\modelcell{UI-TARS-1.5}
  & \heatcellscore{93.0}
  & \heatcellscore{91.9}
  & \heatcellscore{90.5}
  & \heatcellscore{79.4} \\
\midrule
\textbf{All (weighted)}
  & \heatcellscore{80.7}
  & \heatcellscore{76.6}
  & \heatcellscore{77.6}
  & \heatcellscore{74.8} \\
\bottomrule
\end{tabular}
\end{table}

\begin{table}[t]
\centering
\small  
\caption{\textbf{Agent identifiability at both the model and family level.}
Macro F1 for agent identification across 2Wiki, FRAMES, WebShop, and DeepShop.
\textit{Shaded header rows} show family-level classification (10-way); indented
rows show individual-model classification (14-way). Each cell reports the best
score across classifiers (RF, XGB, LSTM, Lasso, LR), with heat shading
encoding value. Bottom row: weighted-average F1 (best classifier: XGB in all).}
\label{tab:combined}
\setlength{\tabcolsep}{5pt}
\begin{tabular}{@{}L{3.55cm}*{4}{C{2.3cm}}@{}}
\toprule
\textbf{\normalsize Model} &
\textbf{\normalsize 2Wiki} &
\textbf{\normalsize FRAMES} &
\textbf{\normalsize WebShop} &
\textbf{\normalsize DeepShop} \\
\midrule

\rowcolor{familybg}
\modelcell{\textbf{Seed-2} \textnormal{\textit{(family)}}}
  & \heatcellscore{95.3}
  & \heatcellscore{95.9}
  & \heatcellscore{93.9}
  & \heatcellscore{87.2} \\
\modelcell{\hspace{0.6em}Seed-2.0-Lite}
  & \heatcellscore{96.1}
  & \heatcellscore{96.7}
  & \heatcellscore{96.1}
  & \heatcellscore{90.2} \\[2pt]

\rowcolor{familybg}
\modelcell{\textbf{GPT-5} \textnormal{\textit{(family)}}}
  & \heatcellscore{90.2}
  & \heatcellscore{80.8}
  & \heatcellscore{74.4}
  & \heatcellscore{69.1} \\
\modelcell{\hspace{0.6em}GPT-5.4}
  & \heatcellscore{91.5}
  & \heatcellscore{78.5}
  & \heatcellscore{68.8}
  & \heatcellscore{68.3} \\[2pt]

\rowcolor{familybg}
\modelcell{\textbf{Claude 4} \textnormal{\textit{(family)}}}
  & \heatcellscore{64.4}
  & \heatcellscore{65.8}
  & \heatcellscore{67.2}
  & \heatcellscore{73.5} \\
\modelcell{\hspace{0.6em}Claude Opus 4.6}
  & \heatcellscore{70.5}
  & \heatcellscore{68.7}
  & \heatcellscore{67.1}
  & \heatcellscore{75.0} \\[2pt]

\rowcolor{familybg}
\modelcell{\textbf{Gemini-3} \textnormal{\textit{(family)}}}
  & \heatcellscore{68.2}
  & \heatcellscore{72.5}
  & \heatcellscore{75.2}
  & \heatcellscore{60.9} \\
\modelcell{\hspace{0.6em}Gemini-3.1-Pro}
  & \heatcellscore{65.2}
  & \heatcellscore{75.2}
  & \heatcellscore{75.2}
  & \heatcellscore{61.1} \\[2pt]

\rowcolor{familybg}
\modelcell{\textbf{Gemini-3-Flash} \textnormal{\textit{(family)}}}
  & \heatcellscore{77.2}
  & \heatcellscore{52.4}
  & \heatcellscore{71.8}
  & \heatcellscore{68.5} \\
\modelcell{\hspace{0.6em}Gemini-3-Flash}
  & \heatcellscore{76.1}
  & \heatcellscore{54.0}
  & \heatcellscore{71.1}
  & \heatcellscore{74.3} \\[2pt]

\rowcolor{familybg}
\modelcell{\textbf{Gemma-4} \textnormal{\textit{(family)}}}
  & \heatcellscore{86.5}
  & \heatcellscore{77.6}
  & \heatcellscore{76.7}
  & \heatcellscore{69.9} \\
\modelcell{\hspace{0.6em}Gemma-4-31B-it}
  & \heatcellscore{82.1}
  & \heatcellscore{84.8}
  & \heatcellscore{74.2}
  & \heatcellscore{63.4} \\
\modelcell{\hspace{0.6em}Gemma-4-26B}
  & \heatcellscore{72.9}
  & \heatcellscore{64.4}
  & \heatcellscore{72.3}
  & \heatcellscore{57.6} \\[2pt]

\rowcolor{familybg}
\modelcell{\textbf{GLM-4.6V} \textnormal{\textit{(family)}}}
  & \heatcellscore{68.4}
  & \heatcellscore{75.9}
  & \heatcellscore{67.1}
  & \heatcellscore{71.0} \\
\modelcell{\hspace{0.6em}GLM-4.6V}
  & \heatcellscore{64.6}
  & \heatcellscore{61.5}
  & \heatcellscore{80.8}
  & \heatcellscore{75.6} \\
\modelcell{\hspace{0.6em}GLM-4.6V-Flash}
  & \heatcellscore{81.9}
  & \heatcellscore{93.2}
  & \heatcellscore{54.1}
  & \heatcellscore{72.7} \\[2pt]

\rowcolor{familybg}
\modelcell{\textbf{Qwen3-VL} \textnormal{\textit{(family)}}}
  & \heatcellscore{90.7}
  & \heatcellscore{94.7}
  & \heatcellscore{91.3}
  & \heatcellscore{87.3} \\
\modelcell{\hspace{0.6em}Qwen3-VL-30B}
  & \heatcellscore{92.7}
  & \heatcellscore{81.8}
  & \heatcellscore{73.5}
  & \heatcellscore{83.8} \\
\modelcell{\hspace{0.6em}Qwen3-VL-8B}
  & \heatcellscore{82.6}
  & \heatcellscore{83.5}
  & \heatcellscore{75.2}
  & \heatcellscore{68.4} \\[2pt]

\rowcolor{familybg}
\modelcell{\textbf{Qwen3.5} \textnormal{\textit{(family)}}}
  & \heatcellscore{75.4}
  & \heatcellscore{76.6}
  & \heatcellscore{79.6}
  & \heatcellscore{76.4} \\
\modelcell{\hspace{0.6em}Qwen3.5-27B}
  & \heatcellscore{90.9}
  & \heatcellscore{98.7}
  & \heatcellscore{82.2}
  & \heatcellscore{87.7} \\
\modelcell{\hspace{0.6em}Qwen3.5-9B}
  & \heatcellscore{63.7}
  & \heatcellscore{65.2}
  & \heatcellscore{74.1}
  & \heatcellscore{64.9} \\[2pt]

\rowcolor{familybg}
\modelcell{\textbf{UI-TARS-1.5} \textnormal{\textit{(family)}}}
  & \heatcellscore{93.0}
  & \heatcellscore{91.9}
  & \heatcellscore{90.5}
  & \heatcellscore{79.4} \\
\modelcell{\hspace{0.6em}UI-TARS-1.5-7B}
  & \heatcellscore{91.3}
  & \heatcellscore{90.1}
  & \heatcellscore{92.1}
  & \heatcellscore{82.9} \\

\midrule
\rowcolor{familybg}
\textbf{All -- family (weighted)}
  & \heatcellscore{80.7}
  & \heatcellscore{76.6}
  & \heatcellscore{77.6}
  & \heatcellscore{74.8} \\
\textbf{All -- model (weighted)}
  & \heatcellscore{79.4}
  & \heatcellscore{75.3}
  & \heatcellscore{74.2}
  & \heatcellscore{72.6} \\
\bottomrule
\end{tabular}
\end{table}

\clearpage

\subsection{Confidence Intervals for Closed-Set Results}
\label{app:bootstrap}
\begin{figure}[h!]
    \centering
    \includegraphics[width=\linewidth]{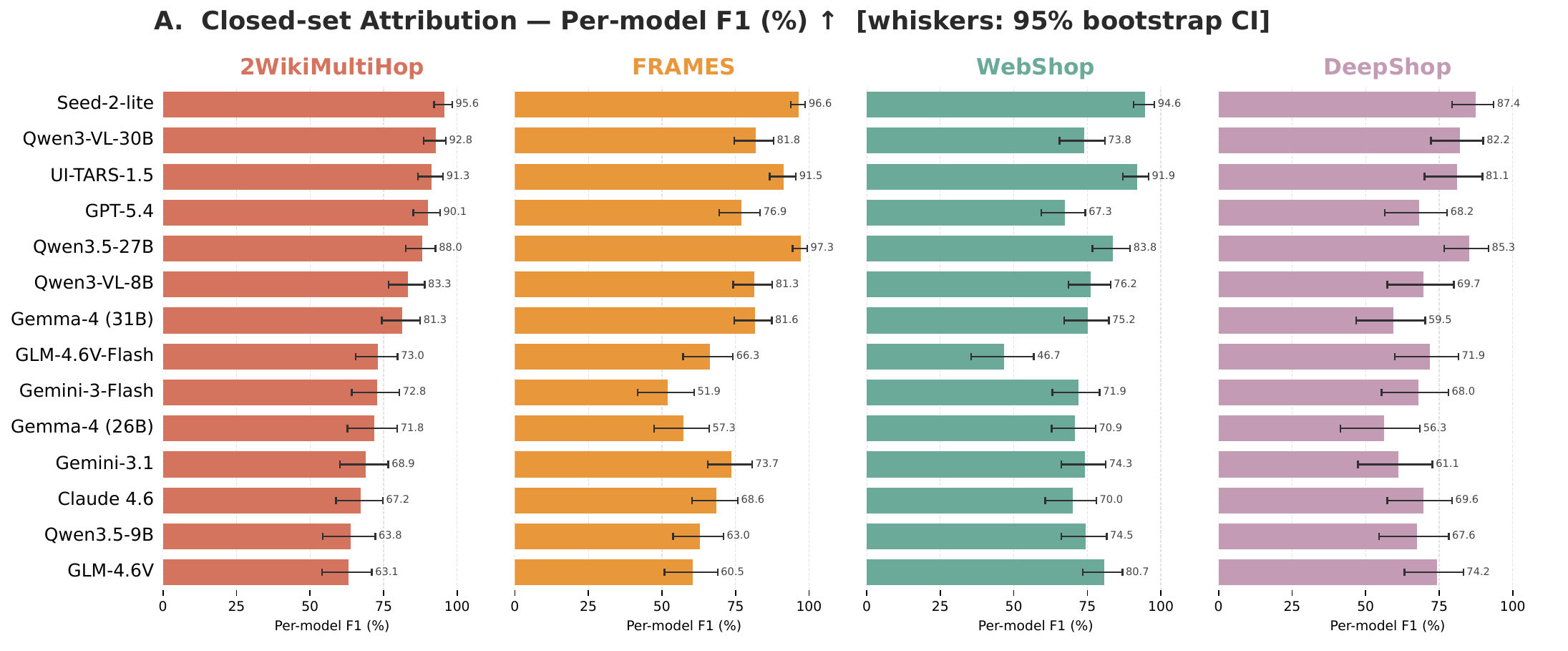}
    \caption{\textbf{Closed-set attribution with bootstrap confidence intervals.} Per-model F1 (\%) with 95\% bootstrap CIs over test traces; bars show the bootstrap mean.}
    \label{fig:bootstrap}
\end{figure}

Figure~\ref{fig:bootstrap} reports per-model closed-set F1 with 95\% bootstrap confidence intervals, obtained by resampling test traces with replacement 500 times and taking the percentile interval. Bars show the bootstrap mean, which can differ slightly from the single-run point estimates in Figure~\ref{fig:main_closed_bars} because of resampling on small test sets. Intervals are widest on DeepShop (38 test episodes).

\clearpage

\subsection{Cost of Updating the Classifier as the Candidate Set Grows}
\label{app:scaling}
We retrain the closed-set XGBoost classifier from scratch as the candidate set grows from 2 to 14 models, using MidScene.js traces from 2WikiMultiHopQA and WebShop and five seeds. Table~\ref{app:tab:scaling} reports Macro-F1 and median fitting time at each candidate set size.


\begin{table}[h!]
\centering
\caption{\textbf{The classifier remains fast to retrain as the number of candidate models grows.}
XGBoost is retrained from scratch on MidScene.js traces for the first $k$ models.
We report Macro-F1 (mean $\pm$ SD over five seeds) and median fitting time in seconds.
Fitting time rises roughly linearly with $k$, reaching about 3\,s at $k{=}14$, and median model size grows from about 0.2\,MiB at $k{=}2$ to about 5.8\,MiB at $k{=}14$.}
\label{app:tab:scaling}
\small
\begin{tabular}{rc cc cc}
\toprule
& & \multicolumn{2}{c}{Macro-F1} & \multicolumn{2}{c}{Fit time (s)} \\
\cmidrule(lr){3-4} \cmidrule(lr){5-6}
$k$ & Chance ($1/k$) & 2WikiMultiHopQA & WebShop & 2WikiMultiHopQA & WebShop \\
\midrule
2  & 0.500 & $0.968 \pm 0.024$ & $0.963 \pm 0.029$ & 0.23 & 0.25 \\
4  & 0.250 & $0.919 \pm 0.051$ & $0.900 \pm 0.053$ & 0.72 & 0.71 \\
6  & 0.167 & $0.883 \pm 0.031$ & $0.849 \pm 0.070$ & 1.13 & 1.16 \\
8  & 0.125 & $0.851 \pm 0.031$ & $0.807 \pm 0.045$ & 1.60 & 1.64 \\
10 & 0.100 & $0.807 \pm 0.024$ & $0.791 \pm 0.030$ & 2.10 & 2.13 \\
12 & 0.083 & $0.793 \pm 0.030$ & $0.760 \pm 0.021$ & 2.57 & 2.62 \\
14 & 0.071 & $0.779 \pm 0.007$ & $0.735 \pm 0.004$ & 3.01 & 3.13 \\
\bottomrule
\end{tabular}
\end{table}

Macro-F1 decreases as more candidates are added, from 0.968 and 0.963 at $k{=}2$ to 0.779 and 0.735 at $k{=}14$, but remains far above chance ($1/k$). Fitting time grows roughly linearly, reaching about 3\,s at $k{=}14$.

\clearpage

\subsection{Does Task Capability Predict Identifiability?}
\label{sec:capability-identifiability}
A central premise of our work is that each agent produces
behaviourally distinct traces that are identifiable. A natural
question is whether these traces are simply a proxy for task
capability: if agents of similar capability tend to generate
similar behavioural traces, our classifiers would be grouping
agents by capability rather than recognising individual identity,
and we would expect a significant correlation between task accuracy
and identifiability. We explore this by examining whether such a
correlation exists.

We measure capability as each agent's accuracy on FRAMES\citep{krishna_fact_2025}. Each agent is given a maximum
of 40 turns to attempt each of the 75 questions in the test set.
We evaluate the agent's final response using an LLM-as-a-judge
setup with \texttt{gpt-5.4-mini}, following the procedure described
by \citet{krishna_fact_2025}; the evaluation prompt is provided below. Questions for which an agent failed
to return a response within 40 turns are marked as incorrect;
Table~\ref{tab:completion-counts} reports completion counts per
agent. We pair each agent's accuracy with its identifiability,
defined as the macro-averaged $F_1$ of the best-performing
classifier (XGBoost) in the closed-set fingerprinting setting, and compute both
Pearson's $r$ and Spearman's $\rho$ across all 14 agents.

\begin{figure}[H]
\centering
\includegraphics[width=1.0\linewidth]{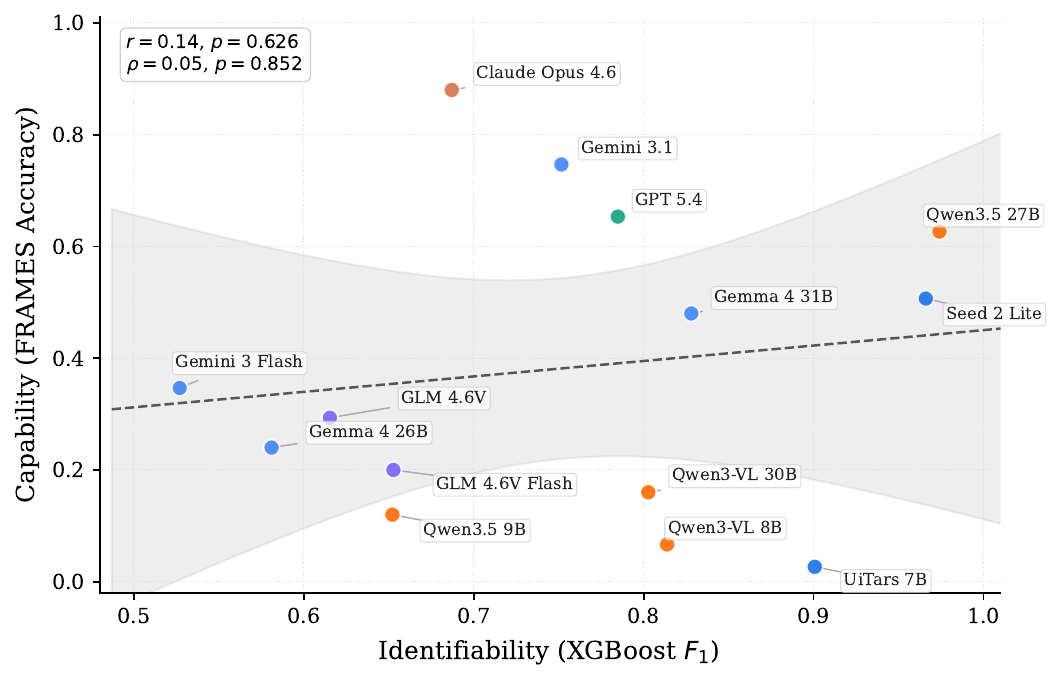}
\caption{\textbf{Task capability does not predict agent identifiability.} Each point represents one of the 14 agents, with identifiability (XGBoost macro $F_1$ in the closed-set setting) on the $x$-axis and task capability (accuracy on FRAMES) on the $y$-axis. The dashed line shows the linear regression fit with its 95\% confidence band. Neither Pearson's $r=0.14$ ($p=0.626$) nor Spearman's $\rho=0.05$ ($p=0.852$) is statistically significant, indicating that the behavioural signatures that make an agent identifiable are largely orthogonal to its task capability.}
\label{fig:capability-identifiability}
\end{figure}

Figure~\ref{fig:capability-identifiability} shows no meaningful
relationship between the capability and identifiability. Neither correlation reaches
statistical significance: Pearson's $r = 0.14$ ($p = 0.626$) and
Spearman's $\rho = 0.05$ ($p = 0.852$). Agents span the full range
of identifiability regardless of their accuracy: Claude Opus~4.6
achieves the highest accuracy (0.88) with moderate identifiability
($F_1 = 0.69$), while UITars-7B is among the most identifiable
agents ($F_1 = 0.90$) yet records the lowest accuracy (0.03).

This supports the view that each agent carries a distinctive
behavioural fingerprint, shaped by how it sequences actions,
manages tool calls, and navigates pages, that is orthogonal to
its task performance. Agents do not converge on shared patterns
simply because they perform similarly; rather, behavioural identity
persists across the capability spectrum.

\begin{center}
\small
\captionof{table}{Agent completion counts, penalised accuracy, and closed-set identifiability (XGBoost $F_1$) on the FRAMES test set (75 questions, 40-turn limit). Accuracy treats incomplete questions as incorrect. \textbf{Bold} denotes the highest value in each column.}
\label{tab:completion-counts}
\vspace{4pt}
\begin{tabular}{lccc}
\toprule
\textbf{Agent} & \textbf{Completed} & \textbf{Accuracy (\%)} & \textbf{Ident.\ $F_1$ (\%)} \\
\midrule
Claude Opus 4.6      & 72 & \textbf{88.00} & 68.71 \\
Gemini 3.1           & 64 & 74.67 & 75.16 \\
GPT-5.4              & \textbf{75} & 65.33 & 78.48 \\
Qwen3.5-27B          & 54 & 62.67 & \textbf{97.40} \\
Seed 2 Lite          & 50 & 50.67 & 96.60 \\
Gemma-4-31B-it       & 58 & 48.00 & 82.80 \\
Gemini 3 Flash       & 27 & 34.67 & 52.70 \\
GLM-4.6V             & 46 & 29.33 & 61.54 \\
Gemma-4-26B-A4B-it   & 26 & 24.00 & 58.11 \\
GLM-4.6V Flash       & 39 & 20.00 & 65.28 \\
Qwen3-VL-30B-A3B     & 35 & 16.00 & 80.27 \\
Qwen3.5-9B           & 25 & 12.00 & 65.22 \\
Qwen3-VL-8B          & 17 &  6.67 & 81.38 \\
UITars-7B            & 10 &  2.67 & 90.07 \\
\bottomrule
\end{tabular}
\end{center}

\begin{tcolorbox}[
  breakable,
  colback=blue!5,
  colframe=blue!40,
  boxrule=0.5pt,
  arc=2pt,
  boxsep=5pt,
  left=8pt,
  right=8pt,
  top=6pt,
  bottom=6pt,
  title={\small\textbf{Autorating Prompt}}
]
\small
\textbf{===Task===}\\
I need your help in evaluating an answer provided by an LLM against a ground truth answer. Your task is to determine if the ground truth answer is present in the LLM's response. Please analyze the provided data and make a decision.

\vspace{4pt}
\textbf{===Instructions===}
\begin{enumerate}
  \setlength{\itemsep}{2pt}
  \setlength{\parskip}{0pt}
  \setlength{\parsep}{0pt}
  \item Carefully compare the ``Predicted Answer'' with the ``Ground Truth Answer''.
  \item Consider the substance of the answers -- look for equivalent information or correct answers. Do not focus on exact wording unless the exact wording is crucial to the meaning.
  \item Your final decision should be based on whether the meaning and the vital facts of the ``Ground Truth Answer'' are present in the ``Predicted Answer''.
\end{enumerate}

\vspace{4pt}
\textbf{===Input Data===}
\begin{itemize}
  \setlength{\itemsep}{2pt}
  \setlength{\parskip}{0pt}
  \setlength{\parsep}{0pt}
  \item Question: \texttt{\{question\}}
  \item Predicted Answer: \texttt{\{LLM\_response\}}
  \item Ground Truth Answer: \texttt{\{ground\_truth\_answer\}}
\end{itemize}

\vspace{4pt}
\textbf{===Output Format===}\\
Provide your final evaluation in the following format:\\
``Explanation:'' (How you made the decision?)\\
``Decision:'' (``TRUE'' or ``FALSE'')

\vspace{4pt}
Please proceed with the evaluation.
\end{tcolorbox}

\clearpage
\subsection{How well do our classifiers transfer across tasks and websites?}\label{app:site_transfer}
The main paper reports in-domain attribution, where train and test traces come from the same task distribution. Here we ask how far these fingerprints transfer across task and website boundaries. We compare three regimes: cross-task transfer, where a classifier is trained on one task and evaluated on another task hosted on the same website; pooled-site training, where traces from multiple tasks on the same website are combined before testing on each held-out test set; and cross-site transfer, where a classifier trained on one website is evaluated on another.

\begin{table}[h!]
\centering
\caption{\textbf{Training on diverse varied behaviours increases classifier performance.}
We present generalisation results across task and website boundaries.
Single-task transfer across tasks on the same site is substantially weaker than in-domain attribution, but pooling multiple tasks from the same website recovers strong performance. Cross-site transfer remains weak, suggesting that behavioural fingerprints are site-conditioned rather than universal.}
\label{tab:generalisation}
\small
\setlength{\tabcolsep}{4pt}
\resizebox{\linewidth}{!}{%
\begin{tabular}{llllc}
\toprule
\textbf{Train} & \textbf{Test} & \textbf{Website} & \textbf{Setting} & \textbf{Macro F1} \\
\midrule
2Wiki & 2Wiki & Wikipedia & in-domain & 79.4 \\
FRAMES & FRAMES & Wikipedia & in-domain & 75.3 \\
WebShop & WebShop & Amazon & in-domain & 74.3 \\
DeepShop & DeepShop & Amazon & in-domain & 72.6 \\
\midrule
2Wiki & FRAMES & Wikipedia & cross-task & 41.1 \\
FRAMES & 2Wiki & Wikipedia & cross-task & 49.8 \\
2Wiki + FRAMES & 2Wiki test & Wikipedia & pooled-site & 81.3 \\
2Wiki + FRAMES & FRAMES test & Wikipedia & pooled-site & 77.2 \\
\midrule
WebShop & DeepShop & Amazon & cross-benchmark & 52.7 \\
DeepShop & WebShop & Amazon & cross-benchmark & 55.9 \\
WebShop + DeepShop & WebShop test & Amazon & pooled-site & 78.8 \\
WebShop + DeepShop & DeepShop test & Amazon & pooled-site & 70.9 \\
\midrule
Wikipedia pooled & Amazon test & Amazon & cross-site & 29.7 \\
Amazon pooled & Wikipedia test & Wikipedia & cross-site & 26.0 \\
\bottomrule
\end{tabular}}
\end{table}

The results in Table \ref{tab:generalisation} show that behavioural fingerprints are not universal task-invariant signatures. 
Training on one Wikipedia task and testing on the other yields much weaker attribution than in-domain training, with macro F1 dropping from 79.4/75.3 in-domain to 41.1 and 49.8 under cross-task transfer. 
However, pooling 2WikiMultiHopQA and FRAMES training traces recovers strong attribution on both held-out test sets, reaching 81.3 on 2WikiMultiHopQA and 77.2 on FRAMES. 
This suggests that a site operator does not need a fingerprint that transfers from a single task to all possible tasks. Instead, diverse traces collected on the same website are sufficient to learn a robust site-conditioned identifier. 
Cross-site transfer remains much weaker, indicating that the identifying signal is shaped by the interaction between the model, task distribution, harness, and website interface.

\clearpage
\subsection{Which features are important to our classifiers ?}\label{app:feature_importance}

\begin{figure}[h!]
    \centering
    \includegraphics[width=\linewidth]{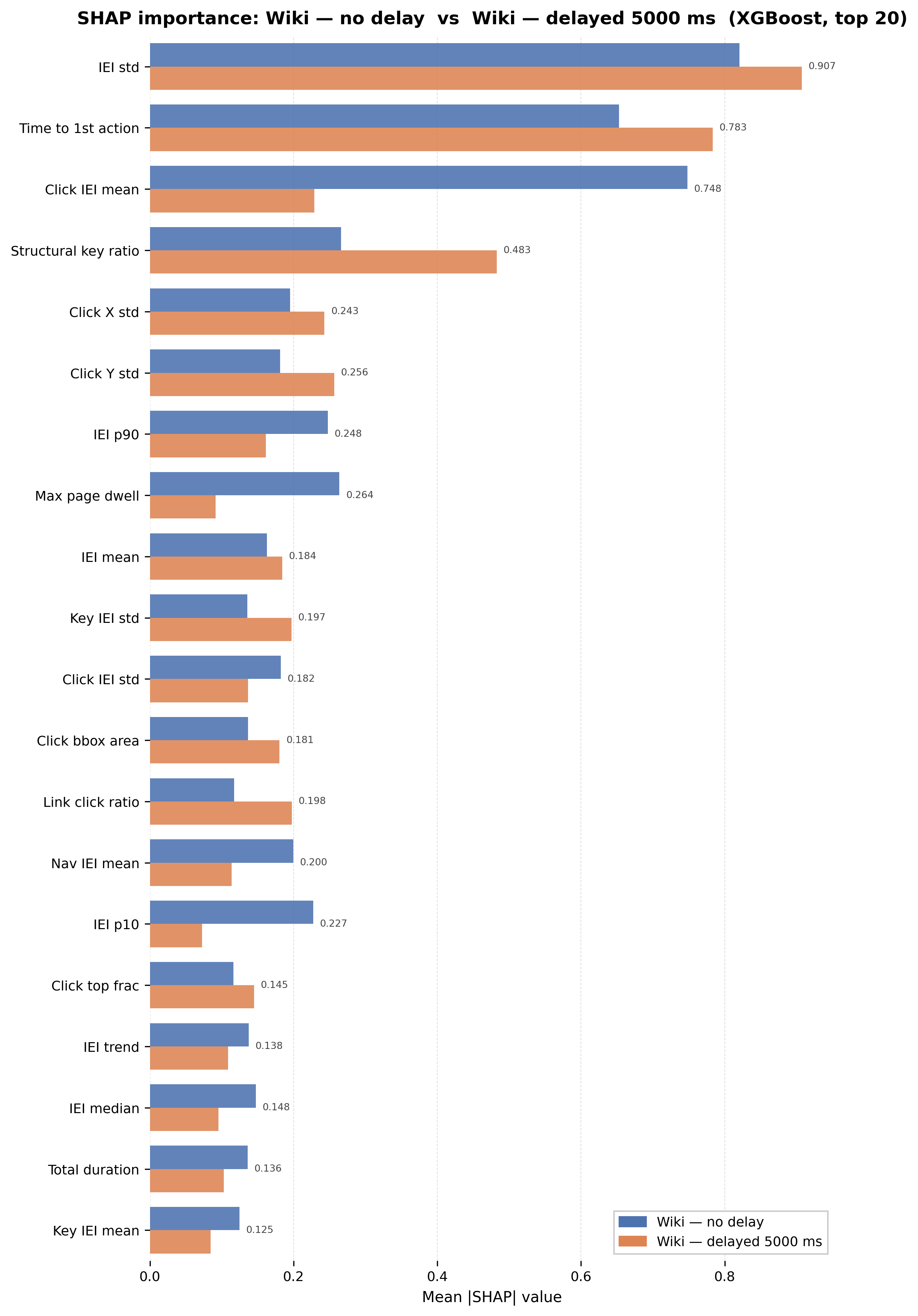}
    \label{fig:placeholder}
\end{figure}

\clearpage
\subsection{Generalisation Across Agent Harnesses}
\label{app:harness-transfer}

Our main experiments train and test our classifiers on traces from a single agent harness, MidScene.js. In deployment, however, a website will be visited by multiple harnesses. To verify that our fingerprinting approach is not specific to MidScene.js and can handle a mixture of harnesses, we also collect traces with the Browser Use harness.

We evaluate six models:
Qwen3.5-27B, GLM-4.6V, Gemma-4-26B-A4B-it, GPT-5.4,
Gemini-3.1, and Claude Opus 4.6.
We use 2WikiMultiHopQA and WebShop and retain the same task-level
train, validation, and test splits used in the main experiments.
We train classifiers on traces collected using MidScene.js, Browser Use,
or a pooled set containing traces from both harnesses.
All results are averaged across five classifier seeds and reported as
mean Macro-F1 $\pm$ standard deviation across seeds.

\begin{table}[h!]
    \centering
    \small
    \setlength{\tabcolsep}{3.5pt}
    \caption{
        Cross-harness model attribution.
        Training and testing within the same harness produces strong
        attribution, whereas zero-shot transfer to an unseen harness falls
        to approximately chance ($1/6 \approx 0.167$).
        Training on both observed harnesses recovers strong performance
        when evaluating on either harness.
    }
    \label{tab:harness-transfer}

    \begin{tabular}{lccc}
        \toprule
        Train harness &
        \multicolumn{3}{c}{Test harness} \\
        \cmidrule(lr){2-4}
        & MidScene.js & Browser Use & Mixed \\
        \midrule
        \multicolumn{4}{l}{\textit{2WikiMultiHopQA}} \\
        MidScene.js
            & $0.812 \pm 0.012$
            & $0.099 \pm 0.004$
            & -- \\
        Browser Use
            & $0.104 \pm 0.007$
            & $0.795 \pm 0.007$
            & -- \\
        Mixed
            & $0.750 \pm 0.010$
            & $0.777 \pm 0.012$
            & $0.758 \pm 0.008$ \\
        \midrule
        \multicolumn{4}{l}{\textit{WebShop}} \\
        MidScene.js
            & $0.827 \pm 0.008$
            & $0.164 \pm 0.013$
            & -- \\
        Browser Use
            & $0.214 \pm 0.014$
            & $0.807 \pm 0.003$
            & -- \\
        Mixed
            & $0.808 \pm 0.009$
            & $0.738 \pm 0.013$
            & $0.771 \pm 0.008$ \\
        \bottomrule
    \end{tabular}
    
\end{table}

As shown in Table~\ref{tab:harness-transfer}, the original within-harness
finding replicates with Browser Use: Macro-F1 ranges from $0.795$ to
$0.827$ when training and evaluation use the same harness. However,
classifiers trained using one harness do not transfer to the other.
Cross-harness Macro-F1 ranges from $0.099$ to $0.214$, approximately
the six-way chance level.

Training on pooled traces from both harnesses recovers most within-harness classifier performance, with Macro-F1 between 
0.738 and  0.808 when evaluated on either harness separately. The fingerprint reflects an interaction between model and harness: the harness fixes what actions are available and how the interaction is structured, and the model decides how to use them. Training on a fixed harness therefore identifies model-specific variation within that setting rather than yielding a harness-invariant classifier.

\clearpage
\subsection{Generalisation Across Inference Engines}
\label{app:engine-transfer}

Because timing features contribute substantially to our fingerprinters, a possible
confound is that the classifier identifies the inference infrastructure
rather than the underlying model. We investigate this using three
open-weight models---GLM-4.6V, GLM-4.6V-Flash, and
Gemma-4-26B-A4B-it---served under both vLLM and SGLang.
All agents use the same Browser Use harness and identical task and browser
configurations.%
We evaluate classifiers using the full feature set, action-timing features
only, and action-content features only. Results are averaged across five
classifier seeds.

Action-timing features describe intervals between browser events and
therefore capture the complete observe--reason--act loop rather than
privileged server-side decoding latency. Action-content features describe
observable behaviour such as action types, click positions, key usage,
scrolling, and navigation structure.

\begin{table}[h!]
    \centering
    \small
    \setlength{\tabcolsep}{3.5pt}
    \caption{
        Model attribution across inference engines.
        Zero-shot transfer between vLLM and SGLang remains strong in the
        tested three-model setting. Action-content features remain
        predictive without timing information, although timing features
        achieve higher Macro-F1.
    }
    \label{tab:engine-transfer}

    \begin{tabular}{lccc}
        \toprule
        Train $\rightarrow$ test &
        Full features &
        Action timing &
        Action content \\
        \midrule
        vLLM $\rightarrow$ SGLang
            & $0.843 \pm 0.005$
            & $0.848 \pm 0.002$
            & $0.702 \pm 0.005$ \\
        SGLang $\rightarrow$ vLLM
            & $0.829 \pm 0.005$
            & $0.802 \pm 0.005$
            & $0.727 \pm 0.008$ \\
        Mixed $\rightarrow$ vLLM
            & $0.968 \pm 0.004$
            & $0.934 \pm 0.008$
            & $0.735 \pm 0.010$ \\
        Mixed $\rightarrow$ SGLang
            & $0.936 \pm 0.007$
            & $0.937 \pm 0.002$
            & $0.706 \pm 0.005$ \\
        \bottomrule
    \end{tabular}

\end{table}

Table~\ref{tab:engine-transfer} shows that zero-shot transfer between
inference engines remains strong: classifiers trained on one engine and
evaluated on the other achieve Macro-F1 between $0.829$ and $0.843$.
Pooling traces from both engines further increases full-feature Macro-F1
to $0.936$--$0.968$.

Timing-only attribution also transfers strongly across the two engines.
Nevertheless, action-content features alone achieve Macro-F1 of
$0.702$--$0.727$, demonstrating that identification is not exclusively
driven by timing. In this controlled subset, the fingerprint is therefore
considerably less sensitive to the inference engine than to the agent
harness. This experiment covers only three models and two serving engines,
however, and does not systematically vary GPU hardware or other deployment
parameters.

\clearpage
\subsection{Prompt Normalisation as a Defence}
\label{app:prompt-normalisation}

In addition to randomising interaction timing, we evaluate whether prescribing
a standard browsing procedure can suppress model-specific behavioural
variation. Our prompt-normalisation defence instructs agents to issue one
concise search, inspect results in order, verify product constraints, avoid
unnecessary revisits, take one action at a time, and stop after selecting the
best product. We evaluate this defence on WebShop using Browser Use and four
vLLM-served models: Gemma-4-26B-A4B-it, GLM-4.6V, Qwen3.5-27B, and
Qwen3-VL-30B-A3B. Results are averaged across five classifier seeds.

We replace the original WebShop instruction with the normalised prompt template
shown in Box~\ref{box:prompt-normalisation}. 

\begin{promptbox}{Prompt-normalisation defence}
\label{box:prompt-normalisation}
\small
\ttfamily

Find the product that best satisfies this request:

\medskip
"\{question\}"
\medskip

Follow this browsing procedure:

\begin{enumerate}
    \setlength{\itemsep}{1pt}
    \setlength{\parskip}{0pt}
    \item Begin with one concise product search.
    \item Inspect relevant results in their displayed order.
    \item Verify the requested constraints on product pages.
    \item Avoid revisiting pages unless it is necessary to verify a constraint.
    \item Take one browser action at a time.
    \item Add only the best matching product or products to the cart, then stop.
\end{enumerate}

Do not create an account, sign in, proceed to checkout, or make a purchase.
\end{promptbox}

\begin{table}[h!]
    \centering
    \small
    \setlength{\tabcolsep}{3.5pt}
    \caption{
        Effect of prompt normalisation on model attribution.
        Normalisation disrupts a classifier trained on the original prompt
        distribution, but performance recovers when the attacker has access
        to traces generated under the normalised policy.
    }
    \label{tab:prompt-normalisation}

    \begin{tabular}{lcc}
        \toprule
        Classifier training data &
        Original test prompt &
        Normalised test prompt \\
        \midrule
        Original prompts
            & $0.917 \pm 0.007$
            & $0.443 \pm 0.029$ \\
        Normalised prompts
            & --
            & $0.955 \pm 0.002$ \\
        50/50 mixed prompts
            & $0.889 \pm 0.009$
            & $0.910 \pm 0.005$ \\
        \bottomrule
    \end{tabular}

\end{table}

As shown in Table~\ref{tab:prompt-normalisation}, prompt normalisation
substantially reduces attribution by a classifier trained on the original
behaviour distribution, lowering Macro-F1 from $0.917$ to $0.443$.
However, this reduction primarily reflects distribution shift rather than the
removal of identifying information. Retraining on normalised traces recovers
Macro-F1 to $0.955$, while training on an equal mixture of the two prompt
regimes achieves $0.889$ and $0.910$ on the original and normalised test
distributions, respectively. Prompt normalisation may therefore increase the
data required by an attacker when the policy is private or changes over time,
but it does not provide robust protection once matching labelled traces are
available. Moreover, this defence relies on model compliance with textual
instructions; we do not evaluate a harness-enforced normaliser that directly
standardises action timing, coordinates, or interaction sequences.

\clearpage
\subsection{Identification versus Observed Events}
\label{app:early-id}
We train XGBoost on full traces and evaluate using only the first $k$ events of each test trace, reporting Macro-F1 averaged over the four datasets. Table~\ref{tab:early-id} complements Figure~\ref{fig:wiki_early_graph}B with absolute event counts. Performance is low for the first 20 events, then rises steadily and reaches 95\% of full-trace Macro-F1 after 100 events.

\begin{table}[h!]
\centering
\caption{\textbf{Identification improves steadily with the number of observed events.}
Macro-F1 of a classifier trained on full traces and evaluated on the first $k$ events of each test trace, averaged over the four datasets. The range is across datasets. Per-model F1 is computed over the 14 models.}
\label{tab:early-id}
\small
\setlength{\tabcolsep}{4.5pt}
\begin{tabular}{rcccccc}
\toprule
Events & \makecell{Mean\\Macro-F1} & Range & \makecell{\% of\\full trace} & \makecell{Median\\per-model F1} & \makecell{Best\\per-model F1} & \makecell{Models with\\F1 $\geq 0.5$} \\
\midrule
5    & 0.12 & 0.07--0.17 & 16\%  & 0.02 & 0.60 & 0.8 / 14 \\
10   & 0.27 & 0.23--0.34 & 36\%  & 0.24 & 0.81 & 2.5 / 14 \\
20   & 0.29 & 0.26--0.34 & 38\%  & 0.26 & 0.71 & 2.0 / 14 \\
30   & 0.41 & 0.21--0.63 & 54\%  & 0.40 & 0.70 & 5.0 / 14 \\
50   & 0.56 & 0.37--0.77 & 74\%  & 0.55 & 0.86 & 8.5 / 14 \\
75   & 0.68 & 0.63--0.78 & 91\%  & 0.68 & 0.89 & 12.5 / 14 \\
100  & 0.72 & 0.69--0.79 & 95\%  & 0.71 & 0.93 & 13.0 / 14 \\
150  & 0.74 & 0.70--0.79 & 98\%  & 0.74 & 0.94 & 13.0 / 14 \\
Full & 0.75 & 0.72--0.79 & 100\% & 0.76 & 0.94 & 13.8 / 14 \\
\bottomrule
\end{tabular}
\end{table}

\newpage
\clearpage
\section*{NeurIPS Paper Checklist}

\begin{enumerate}

\item {\bf Claims}
    \item[] Question: Do the main claims made in the abstract and introduction accurately reflect the paper's contributions and scope?
    \item[] Answer: \answerYes{} 
    \item[] Justification: The abstract and introduction state three claims: (i) agent actions and timing are sufficient to identify the underlying model with up to 96\% F1, (ii) the attack generalises across model sizes and families and supports open-set detection, and (iii) randomised timing delays do not provide robust protection against an adapted classifier. Each claim is supported by experiments in \Cref{sec:results,sec:analysis} and \Cref{fig:main_closed_bars,fig:main_open_bars,fig:feature_importance_shift,fig:timing_delay,fig:wiki_early_graph}, with the 96\% number matching the per-agent F1 reported for Seed-2-lite on 2WikiMultiHopQA.

\item {\bf Limitations}
    \item[] Question: Does the paper discuss the limitations of the work performed by the authors?
    \item[] Answer: \answerYes{} 
    \item[] Justification: Section 8 (Limitations) discusses the dependence on the harness, the two harnesses and two websites tested, the lack of mixed human and agent traffic, the inference engine and hardware not varied beyond one server. In the main text, we acknowledge the weakness of open-set identification and room for improvement. 

\item {\bf Theory assumptions and proofs}
    \item[] Question: For each theoretical result, does the paper provide the full set of assumptions and a complete (and correct) proof?
    \item[] Answer: \answerNA{} 
    \item[] Justification: The paper is empirical and contains no formal theoretical results.

    \item {\bf Experimental result reproducibility}
    \item[] Question: Does the paper fully disclose all the information needed to reproduce the main experimental results of the paper to the extent that it affects the main claims and/or conclusions of the paper (regardless of whether the code and data are provided or not)?
    \item[] Answer: \answerYes{} 
    \item[] Justification: \Cref{sec:experimentalSetup} specifies the four datasets and their splits (\Cref{tab:Dataset_Breakdown}), the 14 evaluated models with citations (\Cref{app:sec:model_details}), the agent harnesses (Midscene.js, Browser Use) in Apppendix \ref{app:harness-transfer}, the trace collection procedure (injected JS observer, fresh browser per session, millisecond timestamps), the five classifier families with hyperparameters (Appendix~\ref{app:classifier_hyperparams}), and the closed-set and open-set evaluation protocols. Behavioural feature definitions are given in Appendix~\ref{app:behavioral_features}. We additionally release the trace corpus and \href{https://anonymous.4open.science/r/known_actions-B31B}{harness}.

\item {\bf Open access to data and code}
    \item[] Question: Does the paper provide open access to the data and code, with sufficient instructions to faithfully reproduce the main experimental results, as described in supplemental material?
    \item[] Answer: \answerYes{} 
    \item[] Justification: We release our labelled trace corpus across the four web environments and the browser-instrumentation harness, with documentation sufficient to reproduce the closed-set and open-set classification results. A GitHub repository is provided at \nolinkurl{https://github.com/KabakaWilliam/known_actions/}.

\item {\bf Experimental setting/details}
    \item[] Question: Does the paper specify all the training and test details (e.g., data splits, hyperparameters, how they were chosen, type of optimizer) necessary to understand the results?
    \item[] Answer: \answerYes{} 
    \item[] Justification: Train/validation/test splits are reported per-dataset in \Cref{tab:Dataset_Breakdown}. Classifier families (Lasso, Logistic Regression, Random Forest, XGBoost, LSTM) and their hyperparameter settings are described in Appendix~\ref{app:classifier_hyperparams}. Feature definitions are in Appendix~\ref{app:behavioral_features}, and evaluation protocols (per-agent F1, macro F1, leave-one-agent-out AUROC) are specified in \Cref{sec:formulation,sec:experimentalSetup}.

\item {\bf Experiment statistical significance}
    \item[] Question: Does the paper report error bars suitably and correctly defined or other appropriate information about the statistical significance of the experiments?
    \item[] Answer: \answerYes{}{} 
    \item[] Justification: Closed-set per-model F1 is reported with 95\% bootstrap confidence intervals over test traces (Appendix \ref{app:bootstrap}), and the new harness, inference-engine, prompt-normalisation, scaling and open-set experiments report mean and standard deviation over five classifier seeds. The main-text per-model F1 values and the open-set AUROCs for the 14 leave-one-agent-out runs are single-run point estimates. We did not repeat trace collection, as the cost of 14 frontier models across four live environments precluded it.

\item {\bf Experiments compute resources}
    \item[] Question: For each experiment, does the paper provide sufficient information on the computer resources (type of compute workers, memory, time of execution) needed to reproduce the experiments?
    \item[] Answer: \answerYes{} 
    \item[] Justification: \Cref{sec:experimentalSetup} reports that all open-source models were served via vLLM on a node with two NVIDIA H100 GPUs, with closed-source frontier models accessed via their respective APIs and Seed-2-lite via OpenRouter. Classifier training (XGBoost, Random Forest, Lasso, Logistic Regression, LSTM) is lightweight and runs on a single CPU. Total classifier hyperperams sweep are reported in Appendix~\ref{app:classifier_hyperparams}. It takes 3 days to collect all traces for the agents. The total cost of generating the traces for closed-source models is provided in Appendix ~\ref{app:trace_collection_run_configuration}. The additional Browser Use trace collection cost USD 680 and the SG-Lang ablation was about 45 H100-hours. 
    
\item {\bf Code of ethics}
    \item[] Question: Does the research conducted in the paper conform, in every respect, with the NeurIPS Code of Ethics \url{https://neurips.cc/public/EthicsGuidelines}?
    \item[] Answer: \answerYes{} 
    \item[] Justification: The research involves no human subjects, scraped personal data, or copyrighted material beyond standard public benchmarks (2WikiMultiHop, FRAMES, WebShop, DeepShop). Traces were generated by automated agents under our control on public websites used for benchmark evaluation in prior work. We have reviewed the NeurIPS Code of Ethics and conform to it.

\item {\bf Broader impacts}
    \item[] Question: Does the paper discuss both potential positive societal impacts and negative societal impacts of the work performed?
    \item[] Answer: \answerYes{} 
    \item[] Justification: \Cref{sec:implications} (Implications) explicitly characterises the negative impacts of agent fingerprinting: model-specific prompt injection, adversarial cost inflation targeting the user's inference budget, and agent-specific access control including model-conditional content poisoning.
    
\item {\bf Safeguards}
    \item[] Question: Does the paper describe safeguards that have been put in place for responsible release of data or models that have a high risk for misuse (e.g., pre-trained language models, image generators, or scraped datasets)?
    \item[] Answer: \answerNA{} 
    \item[] Justification: The released artefacts are interaction traces from automated agents on public benchmark websites and a browser-instrumentation harness. The traces contain no personally identifying information, no scraped private content, and no exploit code. The classifiers used in the paper are commodity supervised models that any practitioner can retrain from a small number of traces (as described in \Cref{sec:analysis}), so gating the corpus would impose friction on defensive research without materially affecting an adversary's ability to reproduce the attack. We therefore release the corpus and harness openly to support defensive and auditing research.

\item {\bf Licenses for existing assets}
    \item[] Question: Are the creators or original owners of assets (e.g., code, data, models), used in the paper, properly credited and are the license and terms of use explicitly mentioned and properly respected?
    \item[] Answer: \answerYes{} 
    \item[] Justification: All existing assets are cited: the four benchmarks (2WikiMultiHop, FRAMES, WebShop, DeepShop), the 14 LLMs evaluated, and the Midscene.js harness. Models are accessed under their respective terms of use (open-weight licences for locally hosted models, API terms for closed-source models, OpenRouter terms for Seed-2-lite). Benchmark datasets are used in their original form for evaluation, consistent with their published licenses.

\item {\bf New assets}
    \item[] Question: Are new assets introduced in the paper well documented and is the documentation provided alongside the assets?
    \item[] Answer: \answerYes{} 
    \item[] Justification: The corpus of agent interaction traces and the browser-instrumentation harness are documented with a README describing the trace schema, the per-session fields, the JavaScript observer, the per-dataset splits, and example scripts to reproduce the closed-set and open-set classification results.

\item {\bf Crowdsourcing and research with human subjects}
    \item[] Question: For crowdsourcing experiments and research with human subjects, does the paper include the full text of instructions given to participants and screenshots, if applicable, as well as details about compensation (if any)? 
    \item[] Answer: \answerNA{} 
    \item[] Justification: The paper does not involve crowdsourcing or research with human subjects.

\item {\bf Institutional review board (IRB) approvals or equivalent for research with human subjects}
    \item[] Question: Does the paper describe potential risks incurred by study participants, whether such risks were disclosed to the subjects, and whether Institutional Review Board (IRB) approvals (or an equivalent approval/review based on the requirements of your country or institution) were obtained?
    \item[] Answer: \answerNA{} 
    \item[] Justification: The study involves no human subjects.

\item {\bf Declaration of LLM usage}
    \item[] Question: Does the paper describe the usage of LLMs if it is an important, original, or non-standard component of the core methods in this research? Note that if the LLM is used only for writing, editing, or formatting purposes and does \emph{not} impact the core methodology, scientific rigor, or originality of the research, declaration is not required.
    \item[] Answer: \answerYes{} 
    \item[] Justification: LLMs are the subject of study of the paper. The 14 evaluated models, their access modalities (local vLLM, OpenRouter, vendor APIs), and the Midscene.js harness through which they interact with web environments are documented in \Cref{sec:experimentalSetup} and Appendix~\ref{app:sec:model_details}.

\end{enumerate}

\end{document}